%% file: ms.tex
\documentclass[twocolumn,tighten,10pt]{aastex631}
\usepackage{amsmath}
\newcommand{\PMO}{Purple Mountain Observatory, Chinese Academy of Sciences, Nanjing 210023, China}

\begin{document}
\title{A Search for Rotation Measure Flare Candidates in Repeating Fast Radio Bursts}

\author{Ye Li$^*$}
\affiliation{\PMO}
\affiliation{State Key Laboratory of Radio Astronomy and Technology, Purple Mountain Observatory, Chinese Academy of Sciences, Nanjing 210023, China}

\correspondingauthor{Ye Li}
\email{yeli@pmo.ac.cn}

\begin{abstract}
Fast radio bursts (FRBs) are millisecond-duration extragalactic radio transients of unknown origin. Rotation measures (RMs) probe their local magneto-ionic environments and provide important clues to their nature. While RM variability has been observed in several repeating FRBs, it is typically gradual or stochastic. Recently, observations of FRB~20220529 revealed an abrupt RM excursion followed by rapid recovery on week-long timescales, termed an ``RM flare'', suggesting a potentially distinct form of RM variability associated with localized magnetized plasma.
In this work, we perform a systematic search for RM flare candidates in repeating FRBs with multi-epoch RM measurements. Using a $3\sigma$ significance threshold, we identify two candidates with multiple observational epochs (FRB~20121102A and FRB~20201124A) and two additional single-epoch candidates (FRB~20180916B), in addition to the event in FRB~20220529A. Our results suggest that RM flares, if confirmed, may not be rare among repeating FRBs and point to highly dynamic magnetized environments local to the sources. Future high-cadence polarimetric observations, particularly following the discovery of RM excursions, will be essential for confirming these candidates and constraining their physical origin.
\end{abstract}

\section{Introduction}

FRBs are extragalactic radio transients with durations on the order of milliseconds (ms) (\citealt{Lorimer2007,Thornton2013}, see \citealt{zhang2023review} for a review). 
To date, approximately 4000 FRBs have been reported \citep{CHIME2026catalog2}. 
Among them, around 100 are known to repeat \citep{chime-repeaters,chime2023rfrb}, exhibiting burst counts ranging from 2 to over $10^4$ \citep{lidi2021, XuH2022, niuc2022, Zhoudj2022, Zhangyk2023, Zhoudj2025}. Many theoretical models have been proposed to explain FRBs (see \citealt{katz16, platts2019} for reviews), with most models invoking neutron stars or other compact objects (e.g., black holes or white dwarfs). The association of FRB~20200428 with the Galactic magnetar SGR~1935+2154 demonstrates that magnetars can produce low-energy FRB-like bursts \citep{stare22020, chime2020sgr}, although the origin of the broader extragalactic FRB population remains uncertain.

The temporal evolution of repeating FRB properties provides important diagnostics of their progenitors and local environments. Among them, polarization properties, in particular the Faraday rotation measure (RM), provide a sensitive probe of the magneto-ionic environment surrounding FRB sources. RM is proportional to the line-of-sight integral of the electron density and magnetic field, ${\rm RM} \propto \int n_e B_{||} dl$, and observed values span $\sim (1$--$10^5)\ \rm rad\ m^{-2}$. Temporal variations in RM have been reported in several repeating FRBs, revealing diverse behaviors on timescales from seconds to years.

Most previously reported RM variations are gradual or stochastic, and can be interpreted in terms of long-term environmental evolution or turbulent plasma fluctuations \citep{Michilli2018, XuH2022, Anna-Thomas2023}. For example, FRB~20121102A, the first known repeater, showed a long-term decay of RM, from $1.03 \times 10^5~{\rm rad~m^{-2}}$ \citep{Michilli2018} to $2.3 \times 10^4~{\rm rad~m^{-2}}$ over nine years \citep{ZhangJS2026}. This behavior can be explained by the expansion of a young supernova remnant (SNR; \citealt{Piro18}) or a magnetar nebula \citep{margalit18}. Another active FRB with large absolute RM values, FRB~20190520B, exhibited a sign reversal of RM \cite{Anna-Thomas2023}, which can be explained by invoking a binary system or precession. The RM evolution of FRB~20180916B, which shows a 16.5-day periodic activity \citep{chime180916}, exhibits two stable states connected by a secular decrease in the absolute value over about 500 days. FRB~20201124A displayed irregular RM variations, including a sudden disappearance of variability \cite{XuH2022}. Its general trend may be explained by plasma turbulence, although a binary progenitor may also be capable of producing it \cite{wangfy2022}. The binary origin is further supported by the possible periodic RM evolution detected in FRB~20201124A and FRB~20220529A \citep{LiangY2025, xujw2025}.

Recently, observations of FRB~20220529A revealed a different type of RM variability: after a 1.5-year low state of $17\pm 101~\rm rad~m^{-2}$, FRB~20220529A experienced a sudden RM increase to $\sim 2000~{\rm rad~m^{-2}}$, and then dropped back to the low state within 10 days \citep{LiY2026}. This behavior is difficult to reconcile with smooth environmental evolution, such as an expanding supernova remnant, or with stochastic fluctuations, and instead suggests the presence of magnetized plasma crossing the line of sight. The possibility of a filament within a supernova remnant has been discussed \citep{Pandhi2026}. 
A companion star may provide a more straightforward explanation, for example through a large orbital eccentricity or a coronal mass ejection (CME). The origin of this plasma could provide an important diagnostic of the local environment and the nature of FRB~20220529A, and may offer insights into the environments of repeating FRBs more broadly. This behavior, referred to as an ``RM flare'' in the literature, suggests RM flares may represent a potentially distinct phenomenological class of RM variability \citep{LiY2026}.

Despite their potential importance, only one RM flare has been identified to date. It remains unclear whether such events represent a common environmental feature of repeating FRB sources, how frequently such abrupt RM excursions occur, and whether they can be reliably distinguished from general RM variability. Their physical origin is even less well understood. A systematic definition of RM flares, together with a dedicated search for similar events across the broader population of repeating FRBs, is therefore essential for understanding their origin.

In this work, we introduce a phenomenological definition of RM flare candidates based on their statistical significance relative to a source-specific long-term baseline, and apply it to a sample of repeating FRBs with multi-epoch RM measurements. The paper is structured as follows. Section \ref{sec:sample} describes the sample and data, Section \ref{sec:method} presents the candidate selection method, and Section \ref{sec:result} presents the results. We discuss the implications and conclude in Sections \ref{sec:discussion} and \ref{sec:conclusion}.

\section{Sample and Data \label{sec:sample}}

\input{sample}

\subsection{Sample Selection}

Our sample consists of repeating fast radio bursts (FRBs) for which multiple rotation measure (RM) measurements are available across distinct observing epochs. We restrict our analysis to repeating sources in order to establish a temporal baseline and to enable the identification of short-timescale RM excursions relative to longer-term behavior.

FRBs are selected from publicly available catalogs and published literature, including observations from CHIME/FRB and FAST, supplemented by follow-up polarimetric measurements where available. To ensure reliable RM time series, we require at least 10 well-separated bursts with RM measurements, yielding a sample of ten repeating FRBs.
Specifically, the RM measurements of FRB~20121102A are primarily from Arecibo \citep{Michilli2018, Hilmarsson2021}, with additional observations from EVN \citep{Plavin2022}, the Very Large Array (VLA), Effelsberg, and FAST \citep{Wang2025, ZhangJS2026}. The RM measurements of FRB~20180301A are from FAST \citep{Luor2020} and Parkes \citep{kumar2023}. The RM values of FRB~20180916B are mainly from CHIME \citep{Mckinven2023, Ng2025}. \cite{Ng2025} and \cite{Fengy2025} also provide RM measurements for several repeating FRBs, including sources such as FRB~20190208A, FRB~20190303A, and FRB~20190417A.
Owing to its large absolute RM values, observations of FRB~20190520B typically require higher frequencies; accordingly, the RM data are taken from observations with the Green Bank Telescope and Parkes \citep{Anna-Thomas2023}. The RM data for FRB~20201124A are primarily from FAST \citep{XuH2022}, with additional measurements from CHIME \citep{Ng2025}. The RM values of FRB~20220529A are taken from \cite{LiY2026}, while those of FRB~20220912A are from \cite{Zhangyk2023}.

Our sample is not intended to be complete in terms of RM coverage. Instead, it is designed to provide a conservative and well-characterized data set suitable for identifying RM flare candidates under the operational definition introduced in Section~3. Several additional repeating FRBs with multiple RM measurements exist. However, due to limited temporal coverage, it is difficult to assess the presence of RM flare candidates in those sources, and they are therefore not included. As a result, the number of repeating FRBs considered in this work is primarily limited by data quality and temporal coverage rather than by the intrinsic properties of the sources.

\subsection{RM Measurements}

RM values used in this work are taken from published measurements. In the literature, RM values are typically estimated using either RM synthesis or QU-fitting. RM synthesis reconstructs the Faraday depth spectrum, while QU-fitting models the Stokes $Q$ and $U$ spectra as a function of wavelength squared. QU-fitting can provide more consistent estimates for sources with complex or multi-component Faraday structures \citep{Mckinven2021}. Therefore, when both methods are available (e.g., \citealt{Mckinven2023}), we adopt the QU-fitting values for our analysis.

The RM sample used in this work is summarized in Table \ref{tab:rm_summary}. For each repeating FRB, we list the observing instruments, the number of RM measurements ($N_{\rm RM}$), the number of observing epochs ($N_{\rm epoch}$, defined as the number of days with at least one RM measurement), the temporal coverage (MJD range), and the corresponding references. To characterize the distribution of RM values, we also provide the median and standard deviation $\sigma_{\rm RM}$ computed directly from the individual measurements.

\section{RM Flare Candidate Selection}
\label{sec:method}

\begin{figure}[!htb]
    \centering
    \includegraphics[width=\linewidth]{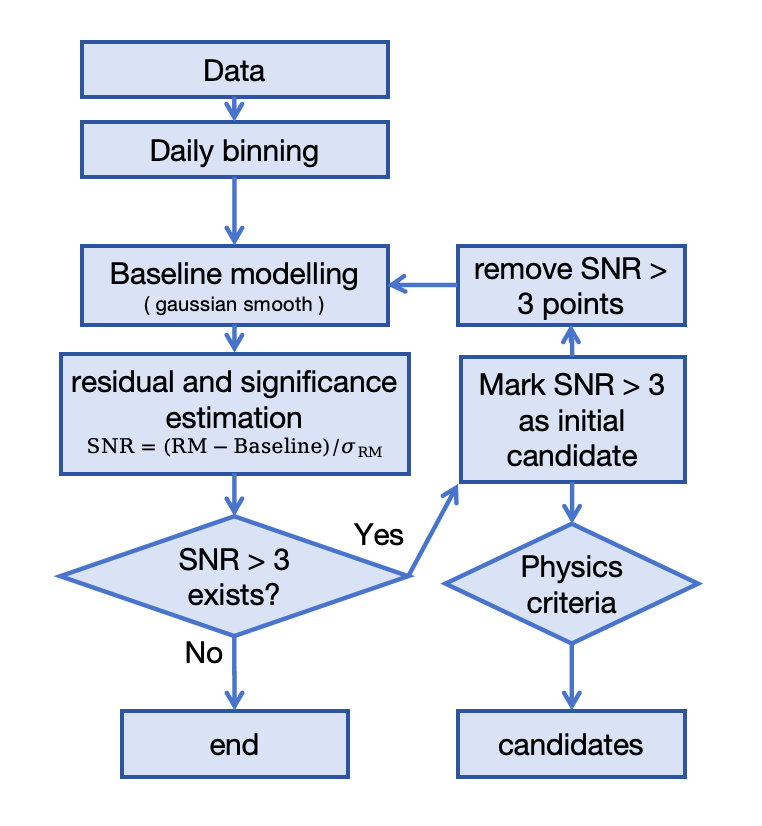}
    \caption{Flowchart of ``RM flare'' candidates search.}
    \label{fig:flowchart}
\end{figure}

The goal of this work is to identify significant deviations in RM on day-to-week timescales of repeating fast radio bursts (FRBs), which we refer to as ``RM flares.'' 
Given that many repeating FRB sources exhibit large and variable RMs with different long-term trends \citep[e.g.,]{Michilli2018,Hilmarsson2021,Mckinven2023,Anna-Thomas2023}, it is necessary to define a source-specific baseline RM evolution before searching for flare-like excursions. The general flowchart of our method is presented in Figure~\ref{fig:flowchart} and described in detail below.

\subsection{Daily Binning}

The sampling of RM measurements is typically governed by the FRB burst rate, resulting in irregular sampling across different days. In addition, RM measurements of repeating FRBs often exhibit significant intra-day fluctuations, which may arise from plasma turbulence. To reduce this effect and obtain a more stable time series, we apply daily binning to the RM data. Following \cite{LiY2026}, we adopt the weighted mean method, where the daily RM value is computed as the weighted average of individual measurements within a day, with weights $w_i = 1 / \epsilon_i^2$, where $\epsilon_i$ is the measurement uncertainty of each detected RM. To account for the intra-day scatter, the daily RM uncertainty is defined as
\begin{equation}
\sigma_{\rm RM}^2 = \frac{1}{\Sigma(1 / \epsilon_i^2)} + {\rm RMS}^2_{\rm RM},
\end{equation}
where $\Sigma$ sums within one day, and ${\rm RMS}_{\rm RM}$ is the standard deviation of those values. The second term accounts for intra-day scattering beyond the measurement uncertainties. The median and standard deviation of the daily-binned RM values for each source are listed in Table~\ref{tab:rm_summary}. These daily-binned statistics are less sensitive to uneven intra-day sampling and burst rate variations, thus providing a more robust characterization of the long-term RM behavior.

\subsection{Baseline Estimation}

After daily binning, we model the baseline RM evolution $\mathrm{RM}_{\rm base}(t)$ for each FRB source. Since the physical origin of RM variations is still unknown, a non-parametric method is preferred. We adopt Gaussian smoothing with a width of $\sigma = 20$ days, which is well matched to the expected duration of RM flares ($1$--$100$ days). The effect of width will be discussed in Section \ref{sec:discussion_width}.

To obtain a baseline that is not biased by the presence of RM flares themselves, we implement an iterative rejection procedure. Starting with an initial Gaussian smoothing applied to all daily-binned data, we identify points with significance (see Section~\ref{sec:residuals}) exceeding a threshold of $3\sigma$. These points are temporarily excluded, and the smoothing is repeated on the remaining data. This process is repeated until no new points are excluded. This iterative approach is widely used in time-series analysis to reduce the impact of outliers and transient features (e.g., \citealt{Huber1964, Cleveland1979}). The final baseline is then used for the subsequent calculations.

\subsection{Residuals and Significance}
\label{sec:residuals}

The residual RM at each observation is defined as
\begin{equation}
\delta \mathrm{RM}_i = \mathrm{RM}_i - \mathrm{RM}_{\rm base}(t_i).
\end{equation}
To account for both measurement uncertainties and the uncertainty in the baseline itself, we define a significance as
\begin{equation}
{\rm SNR}_i = \frac{|\delta \mathrm{RM}_i|}{\sigma_{\rm eff}} = \frac{|\delta \mathrm{RM}_i|}{\sqrt{\sigma_{\rm RM,i}^2 + \sigma_{\rm base}^2}},
\end{equation}
where $\sigma_{\rm eff}$ is the effective uncertainty, $\sigma_{\rm RM,i}$ is the daily measurement error from Equation~(1) and $\sigma_{\rm base}$ is the estimated uncertainty of the baseline, computed as the standard deviation of the residuals $\delta \mathrm{RM}_i$. Since the sign of RM indicates the direction of the magnetic field, we focus on absolute deviations.

\subsection{Candidate Selection \label{sec:selection}}

An RM flare candidate is identified when it satisfies the following criteria:

\begin{enumerate}
    \item \textbf{Statistical significance:} The normalized deviation must satisfy ${\rm SNR}_i \geq 3$. Given that the number of observing epochs for each source in our sample is less than 50, the expected average number of $3\sigma$ false positives is only 0.13 per source, making this a meaningful threshold for initial selection.
    \item \textbf{Local extremum:} The residual must exhibit a local peak or dip around the candidate epoch, ensuring that the deviation is a discrete flare-like event in RM space, rather than a monotonic increase or decrease.
    \item \textbf{Duration:} The excursion must occur on a timescale much shorter than the baseline evolution. We require the duration of the candidate to be $<100$ days, consistent with the expected timescale of transient magnetized structures (e.g., coronal mass ejections or orbital passages).
\end{enumerate}

Candidates are further evaluated based on their temporal coverage. Events detected across multiple epochs are considered more robust, while single-epoch detections are treated with caution and regarded as marginal unless supported by additional evidence.

Candidates that satisfy both the statistical and physical criteria are then examined individually via Gaussian fitting to confirm their flare nature and to measure event-level properties (e.g., peak amplitude, duration, and significance) in Section \ref{sec:result}.

\section{Results}
\label{sec:result}

\begin{figure*}[!htb]
    \centering
    \includegraphics[width=\linewidth]{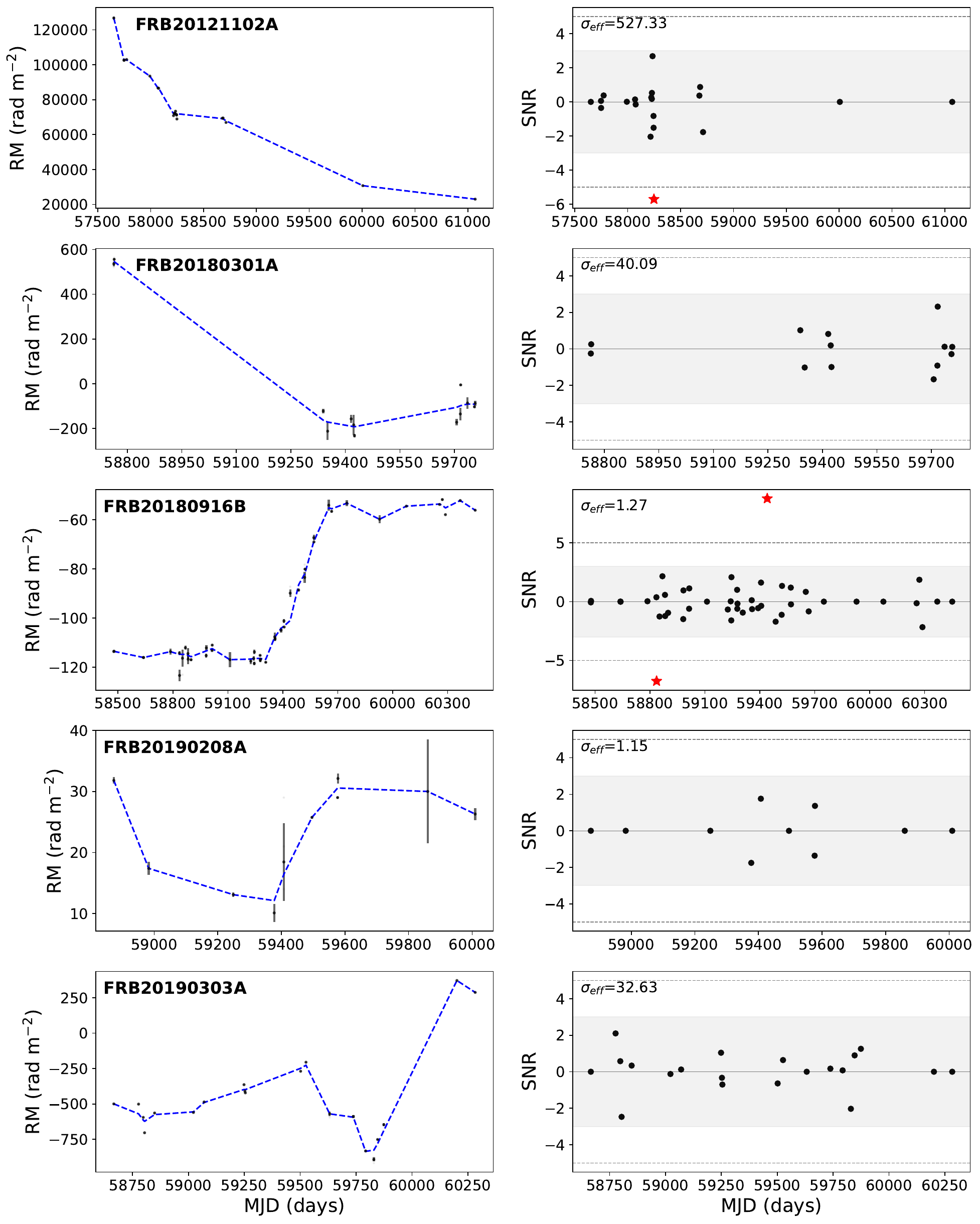}
    \caption{RM flare search for different repeating FRBs. Left panels: RM evolution of each source. Gray dots represent individual measurements, and black points show daily-binned values with uncertainties. The blue dashed lines indicate the estimated baselines from our iterative Gaussian smoothing method. Right panels: Significance (SNR) for each epoch, with the region $-3 < \mathrm{SNR} < 3$ shaded in gray for reference. RM flare candidates with $\mathrm{SNR} > 3$ are marked as red stars.}
    \label{fig:rmflare_candi}
\end{figure*}

\begin{figure*}[!htb]
    \centering
    \includegraphics[width=\linewidth]{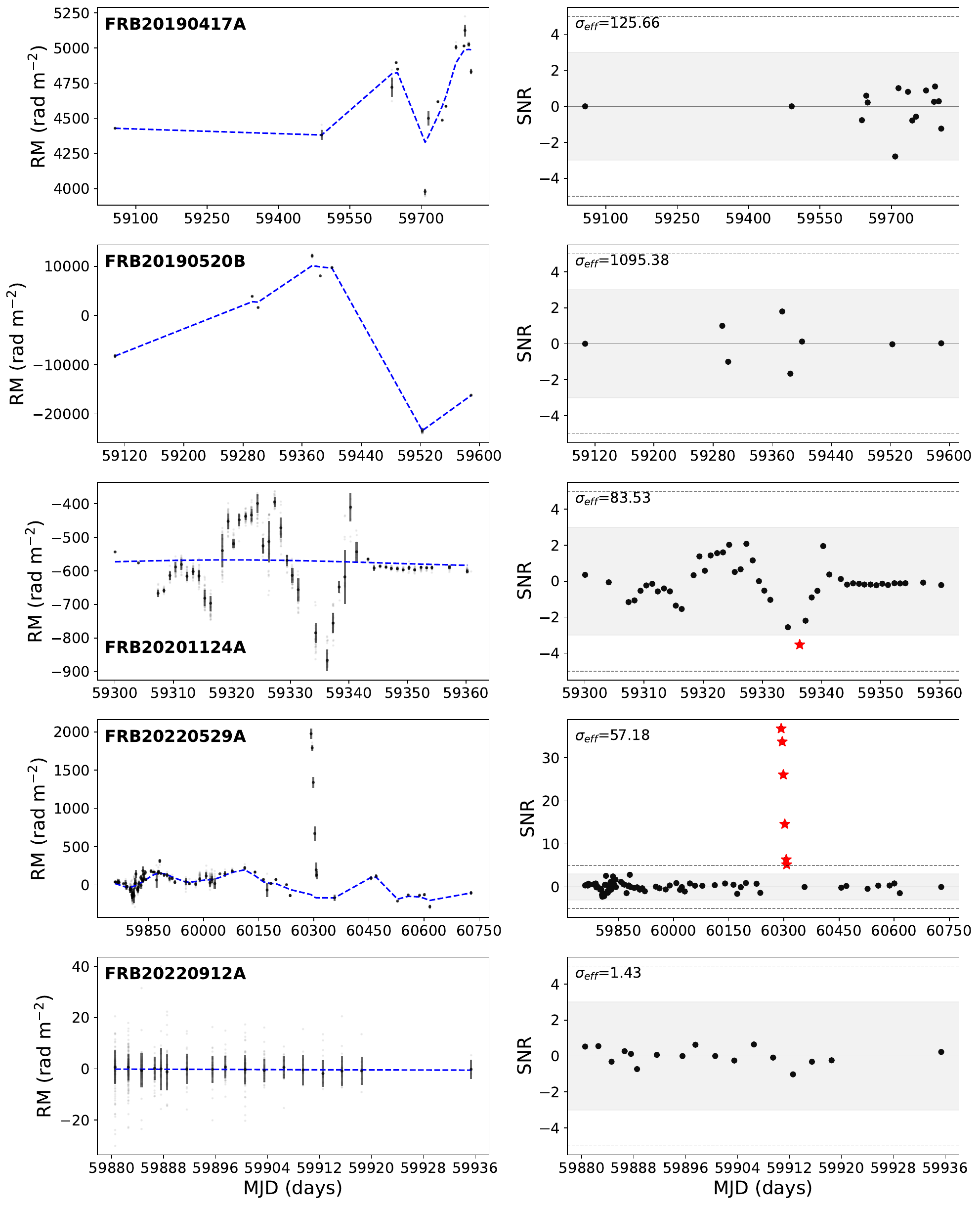}
    \caption{Search results (continued).}
    \label{fig:rmflare_candi_cont}
\end{figure*}

\input{flare_candidates.tex}

\subsection{Initial RM Flare Candidates}

Figure~\ref{fig:rmflare_candi} presents the results of applying the operational definition described in Section~\ref{sec:method} to our sample of repeating FRBs. The left panels show the RM measurements (gray dots) and daily-binned values (black points with error bars). The estimated baselines from the iterative Gaussian smoothing method are shown as blue dashed lines. The right panels display the significance (SNR) for each epoch, with the $\pm3\sigma$ region shaded in gray and candidates with $\mathrm{SNR} > 3$ marked as red stars.

We identify several candidates that satisfy all the criteria outlined in Section~\ref{sec:selection}. These events are characterized by significant RM changes deviating from the long-term RM behavior. The identified candidates are summarized in Table~\ref{tab:flare}, which lists the source name, the MJD, the measured RM, the baseline value, and the corresponding SNR.

Initial candidates are found in FRB~20121102A, FRB~20180916B, FRB~20201124A, and FRB~20220529A. With the exception of FRB 20220529A, the significance of the candidates in the other sources is relatively low, even with the iterative outlier rejection method. This may explain why the first robust RM flare was discovered in FRB~20220529A. Notably, the SNR of the RM flare in FRB~20220529A derived from our method is higher than that estimated in \cite{LiY2026}, which used a pre-flare epoch baseline and low-order polynomial fitting. This demonstrates that our iterative Gaussian smoothing approach improves the efficiency and sensitivity of RM flare searches.

\subsection{RM Time Series Characteristics}

\begin{figure*}[!htb]
    \centering
    \includegraphics[width=0.48\linewidth]{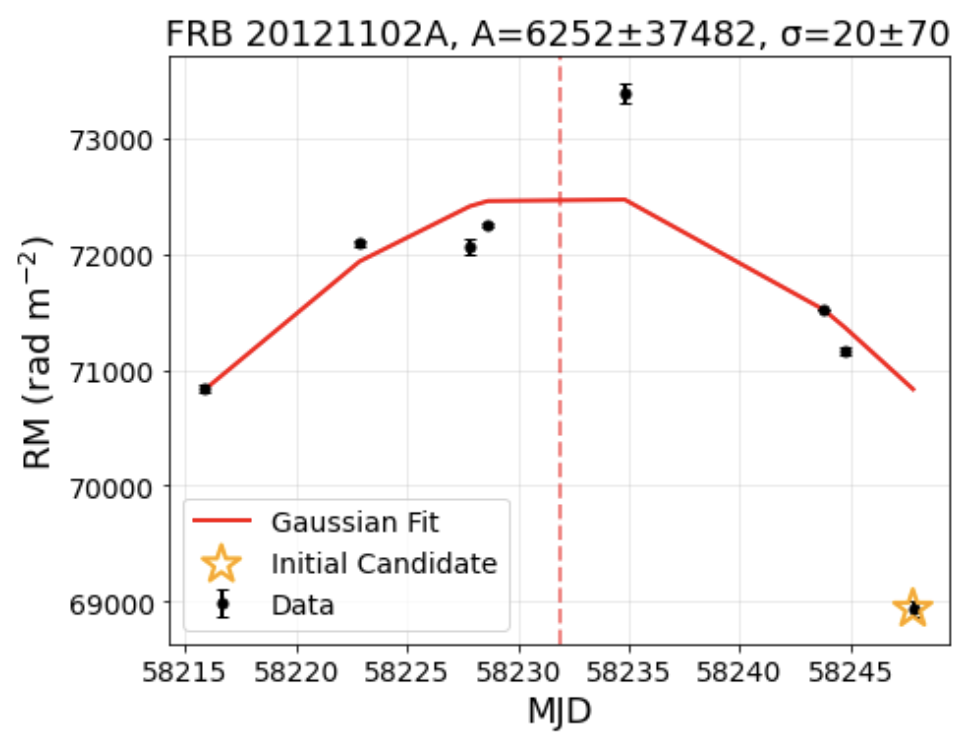}
    \includegraphics[width=0.48\linewidth]{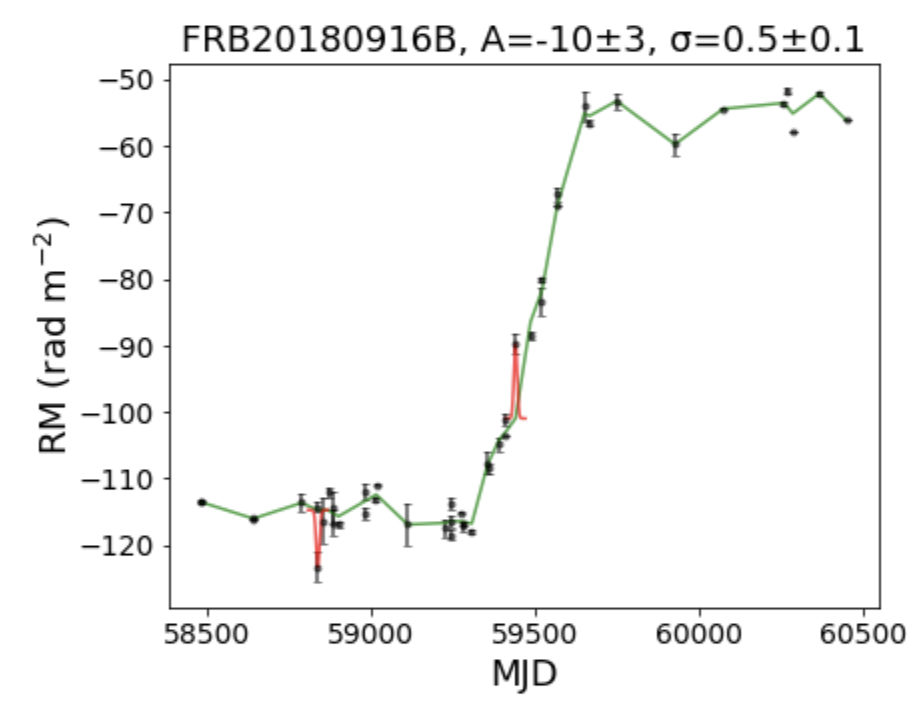}
    \includegraphics[width=0.48\linewidth]{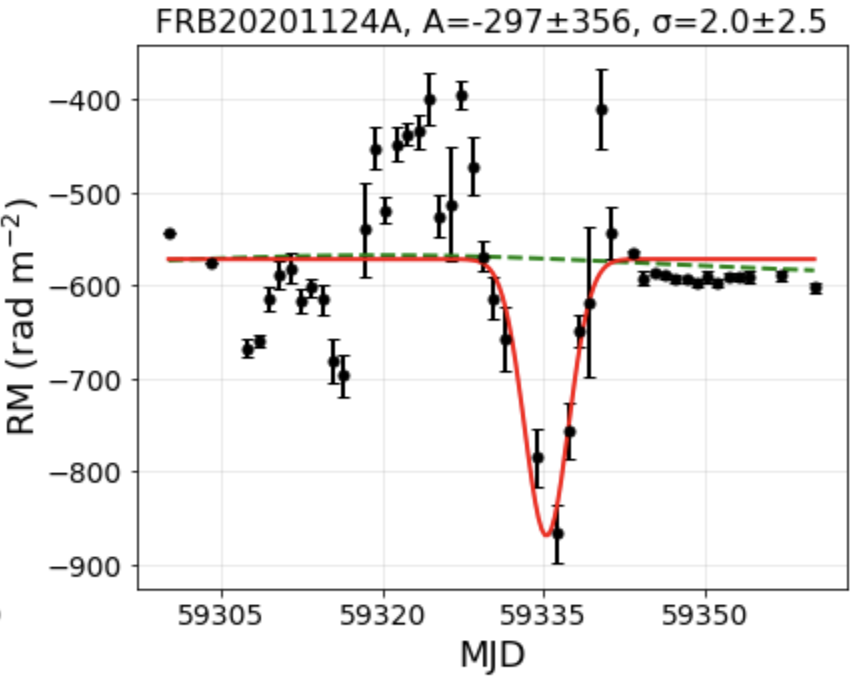}
    \includegraphics[width=0.48\linewidth]{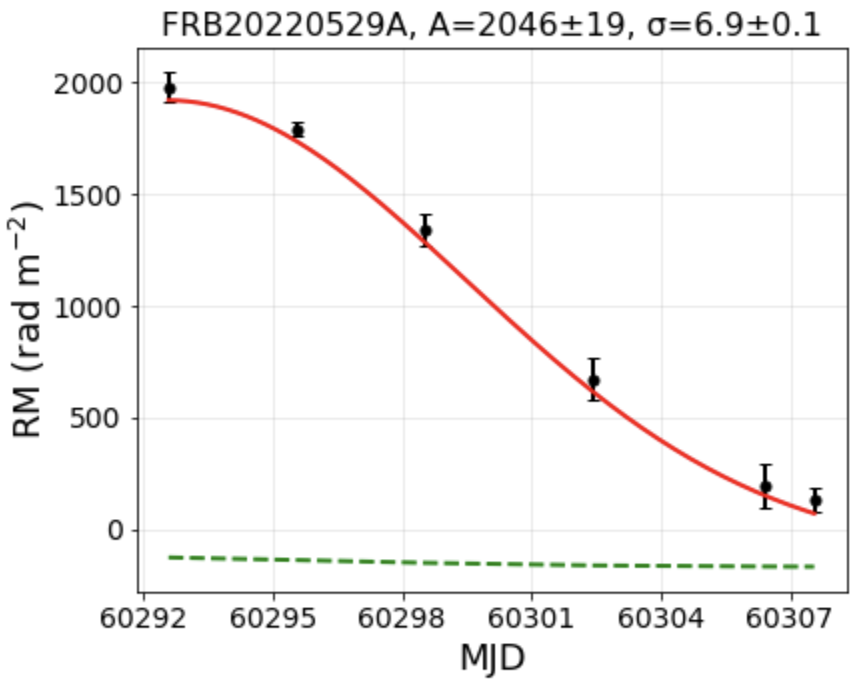}
    \caption{Gaussian fit results of the initial candidates.}
    \label{fig:rmflare_candi}
\end{figure*}

To examine the physical validity of the initial candidates identified in Section~\ref{sec:result}, we further analyze their temporal characteristics. First, we verify the duration and local extremum criteria defined in Section~\ref{sec:selection}. Second, we perform Gaussian fitting ${\rm RM}=A\exp\left(-\frac{(t-{\rm MJD})^2}{2\sigma^2}\right)$ to model the time series of each candidate. This approach is motivated by physical considerations: if an RM flare is associated with orbital motion, the RM evolution can be approximately described by a Gaussian profile \citep{Johnston1996}; similarly, a coronal mass ejection (CME) crossing the line of sight can also produce a Gaussian-like RM variation \citep{Kooi2017}. However, we emphasize that the RM flare evolution is not necessarily Gaussian; the fitting is intended only to characterize the approximate duration, amplitude, and peak time of each event.

\subsubsection{FRB~20220529A}

The most convincing candidate is the one in FRB~20220529A, the prototype of RM flares. It shows a clear, high-SNR excursion that meets all statistical and physical criteria. Thanks to prompt follow-up observations, six epochs of this event are identified as significant. Its Gaussian fit yields a center MJD of 60292, a width of $\sigma_{\rm t} = 6.9 \pm 0.1$ days, and an amplitude of $2046 \pm 19$~rad~m$^{-2}$, on top of a baseline of $-127$~rad~m$^{-2}$. The evolution curve can be well described by a Gaussian profile.

\subsubsection{FRB~20121102A}

The candidate in FRB~20121102A is identified near MJD~58247. However, this point is not an extremum in the RM evolution. Instead, a clear maximum is present at MJD~58234.8. Due to the intermittent sampling of FRB~20121102A, with a typical cadence of more than 100 days, the initial candidate at MJD~58247 may be a spurious identification caused by the maximum at MJD~58234.8, which raises the baseline and makes the later point appear significant. We therefore consider MJD~58234.8 as the true candidate and fit the data around it with a Gaussian. 

The absolute RM of this source shows a secular decreasing trend. Around MJD~58234.8, there are nine bursts across eight epochs with RM detections, enabling a meaningful analysis. These eight epochs yield a Gaussian width of $\sim 20$ days, a median peak time at MJD~58231.9, and an amplitude of $6252$~rad~m$^{-2}$. The evolution is generally Gaussian-like, with a clear peak at MJD~58234.8. However, the baseline of this candidate is difficult to estimate, as there is no stable quiescent state outside these eight epochs. Moreover, the amplitude and width suffer from large uncertainties. We therefore classify it only as a marginal candidate.

\subsubsection{FRB~20201124A}

One initial candidate is identified in FRB~20201124A at MJD~59336.3. A Gaussian fitting reveals an amplitude of $\sim 300$~rad~m$^{-2}$ and a width of $\sim 5$ days, covering seven epochs, although only one epoch has SNR $> 3$. Notably, the overall RM evolution of FRB~20201124A may be interpreted as the signature of a binary system \citep{wangfy2022}. If this is the case, only a small-amplitude RM fluctuation is required to explain the candidate, consistent with our fitting.

\subsubsection{FRB~20180916B}

Two initial candidates are identified in FRB~20180916B, at MJD~58836.2 and MJD~59440.5. However, only a single data point is significant in each case. Furthermore, the candidate at MJD~59440.5 occurs during a rising phase of the RM evolution. Both are therefore considered marginal candidates, and their amplitudes can only be roughly estimated from the residuals between the data points and the baseline.

\subsubsection{Summary}

In summary, besides FRB~20220529A, there are two candidates with multi-epoch detections (FRB~20121102A and FRB~20201124A) and two with only single-epoch detections (FRB~20180916B). 
They can be grouped according to their observational robustness. 
The event in FRB~20220529A represents a robust RM flare with well-sampled temporal evolution and high significance. 
The candidates in FRB~20121102A and FRB~20201124A are considered probable, as they are supported by multiple epochs but remain subject to baseline uncertainties or alternative interpretations. 
The candidates in FRB~20180916B are regarded as marginal due to their single-epoch nature. This grouping reflects observational limitations rather than distinct physical classes.

\section{Discussion \label{sec:discussion}}

\subsection{Occurrence and Selection Effects}

We have identified four additional possible RM flare candidates using a $3\sigma$ threshold besides FRB~20220529A. For a purely Gaussian noise process with 50 independent epochs, the expected number of spurious $3\sigma$ excursions is $\sim 0.13$ per source, or $\sim 1.3$ across our sample of 10 sources. In contrast, we detect a total of 10 candidates. After excluding FRB~20220529A, we are left with 196 epochs from the remaining nine sources, for which the expected number of false detections is $\sim 0.5$. However, four additional candidates remain, still exceeding the expectation from random fluctuations.

This excess suggests that these candidates may not be purely statistical outliers and may indicate a more common mode of RM variability among repeating FRBs than previously recognized. However, given the limited sample size and heterogeneous data quality, this result should be regarded as suggestive rather than definitive.

Several observational biases may have hindered the identification of such events in previous studies. First, many repeating FRBs exhibit large absolute RM values (e.g., FRB~20121102A and FRB~20190520B), which tend to reduce the observed linear polarization fraction and increase RM uncertainties, making both RM detection and variability tracking more challenging for low-frequency instruments such as CHIME and FAST \citep{FengY2022}. Second, the temporal sampling of most repeating FRBs remains sparse ($10$--$100$ days), limiting the ability to capture short-duration RM excursions. Notably, the candidate in FRB~20121102A was detected with Arecibo, the most sensitive single-dish telescope prior to FAST, highlighting the importance of sensitivity for detecting RM variations in faint FRBs. These considerations suggest that high-cadence, high-sensitivity polarimetric monitoring—particularly for sources with relatively low absolute RM values—will be essential for constraining the true occurrence rate of RM flares.

\subsection{Effect of the Width for Gaussian Smoothing}
\label{sec:discussion_width}

We examined the effect of the smoothing width $\sigma_t$ in the iterative Gaussian smoothing method. A width that is too small ($\sigma_t = 1$ day) is sensitive to small-scale fluctuations and fails to recover RM flares with durations of $\sim 10$ days. A width that is too large ($\sigma_t = 100$ days) produces an smooth baseline that reduces the significance of genuine flares and introduces edge effects near the boundaries of the time series.

Figure~\ref{fig:sigmat_comparison} compares the results for $\sigma_t = 1, 10, 20, 30, 100$ days for all sources with RM flare candidates. For $\sigma_t=10, 20, 30$ days, the results are broadly consistent. The candidate near MJD~58230 in FRB~20121102A, the two candidates in FRB~20180916B, and the candidates in FRB~20201124A and FRB~20220529A are all recovered across these widths. The significance varies because the effective uncertainty $\sigma_{\rm eff}$ increases with $\sigma_t$.

When $\sigma_t = 1$ day, even the prominent RM flare in FRB~20220529A becomes undetectable due to the baseline overfitting the data. For $\sigma_t = 100$ days, edge effects become noticeable: the first data point of FRB~20121102A is falsely identified as a candidate. Edge effects are also present for $\sigma_t = 30$ days in FRB~20220529A, caused by the data gap immediately preceding the RM flare, which can occasionally alter the SNR. Based on these tests, we adopt $\sigma_t = 20$ days as the default width for our analysis, as it provides a balance between sensitivity to short-timescale features and stability against noise and edge effects.

\subsection{Effect of Baseline Estimation Methods}
\label{sec:method_comparison}

To examine the impact of different baseline estimation methods, 
we explored a range of approaches and present here three representative and commonly used methods: 
Gaussian process regression with detrending (GPR), polynomial fitting (Poly, order 6), 
and Gaussian smoothing without iteration (GS, width $\sigma_t = 20$ days). These methods span a range of model flexibility, from parametric (polynomial) to non-parametric (GS) and fully probabilistic (GPR).

Figure~\ref{fig:method_comparison} presents the results for the four sources with RM flare candidates: FRB~20121102A, FRB~20180916B, FRB~20201124A, and FRB~20220529A. For nearly monotonic evolution (e.g., FRB~20121102A), all three methods produce broadly consistent baselines. However, polynomial fitting does not adequately capture the more irregular evolution seen in the other three sources. GPR performs well for FRB~20180916B and FRB~20201124A, but tends to fit small-scale structures and does not clearly recover the RM flare in FRB~20220529A. Gaussian smoothing produces results broadly consistent with the iterative Gaussian smoothing adopted in Section~\ref{sec:method}, although the baseline tends to be elevated around the RM flare in FRB~20220529A, reducing its inferred significance.

Most of the candidates listed in Table~\ref{tab:flare} are also identified by at least one of these alternative methods. One notable case is the candidate at MJD~59440 in FRB~20180916B, which has a significance of $\sim 2.5$ with GPR and GS, but exceeds $3\sigma$ when the iterative method is applied.

A few additional spurious candidates are also identified by the alternative methods. For example, polynomial fitting identifies candidates near MJD~58750 in FRB~20121102A and near MJD~59900 in FRB~20180916B, likely due to inadequate baseline modeling. GPR identifies two candidates near MJD~59800 in FRB~20220529A, which are likely related to overfitting of local fluctuations. On the other hand, the candidate near MJD~59340 in FRB~20201124A is identified by both GPR and polynomial fitting, warranting further investigation.

In summary, compared with the iterative Gaussian smoothing method with outlier rejection adopted in Section~\ref{sec:method}, the alternative methods produce broadly consistent baseline estimates and recover similar RM flare candidates. However, GPR can be sensitive to kernel parameter choices and may overfit small-scale structures. Polynomial fitting lacks sufficient flexibility for complex RM evolution, while higher-order polynomials risk overfitting. Gaussian smoothing without outlier rejection may overestimate the baseline in the presence of high-SNR excursions. The iterative Gaussian smoothing with outlier rejection appears to provide a reasonable balance between flexibility and robustness for the sources considered here.

Nevertheless, we caution that RM flare identification remains subject to uncertainties from sparse sampling, incomplete understanding of FRB RM evolution, and possible intrinsic turbulence. More intensive polarimetric monitoring, particularly following the detection of significant RM excursions, will be essential for identifying more robust candidates and constraining their physical origin.

\section{Conclusions}
\label{sec:conclusion}

We have presented a systematic search for rotation measure (RM) flare candidates in repeating fast radio bursts. Our main results are summarized as follows:

\begin{enumerate}
    \item We employ an iterative Gaussian smoothing method with outlier rejection to estimate the RM baseline and identify flare candidates.
    
    \item Applying this method to a sample of repeating FRBs with multi-epoch RM measurements, we identify two tentative candidates with multi-epoch coverage (FRB~20121102A and FRB~20201124A), one marginal candidate with only single-epoch detection (FRB~20180916B), in addition to the well-sampled event in FRB~20220529A.
    
    \item If confirmed, these tentative RM flare candidates suggest that rapidly evolving magnetized environments may be common in at least a subset of repeating FRB sources.
    
    \item The amplitudes of the candidate events range from $\sim10$ to $\sim6000$~rad~m$^{-2}$, and their $1\sigma$ durations range from $\sim0.5$ to $\sim20$ days, although both measurements are subject to significant uncertainties due to sparse sampling.
\end{enumerate}

These results indicate that some repeating FRBs may reside in dynamically evolving magneto-ionic environments. However, we emphasize that the candidates presented here are tentative and require further confirmation. Robust physical interpretation will require higher-cadence and multi-wavelength observations. Future high-cadence polarimetric monitoring will be essential to confirm these candidates, identify more convincing events, assess the prevalence of RM flares, and ultimately constrain their physical origin.

\section*{Acknowledgements}
%\RE{We thank the anonymous referee for helpful suggestions and comments.} 
We thank Qiang Yuan, Songbo Zhang, Xuan Yang, Yi-Fang Liang, Bing Zhang, Yuan-Pei Yang, and the AI assistants (DeepSeek, ChatGPT) for helpful discussions.
This work is supported by the National Natural Science Foundation of China (No. 12393813, 12103089), National Key R\&D Program of China(2024YFA1611700), the Strategic Priority Research Program of the Chinese Academy of Sciences (grant No. XDB0550400), the CAS Project for Young Scientists in Basic Research (Grant No. YSBR-063), the China Manned Space Program with grant No. CMS-CSST-2025-A17, the Natural Science Foundation of Jiangsu Province (Grant No. BK20211000).

%This work is partially supported by the Natural Science Foundation of China (Grant Nos. 12321003, 12041306, 12103089,12393813), the National Key Research and Development Program of China (2022SKA0130100), National Key R\&D Program of China(2024YFA1611704), the Natural Science Foundation of Jiangsu Province (Grant No. BK20211000), International Partnership Program of Chinese Academy of Sciences for Grand Challenges (114332KYSB20210018), the CAS Project for Young Scientists in Basic Research (Grant No. YSBR-063), the CAS Organizational Scientific Research Platform for National Major Scientific and Technological Infrastructure: Cosmic Transients with FAST.

\software{Astropy~\citep{astropy:2013, astropy:2018, astropy:2022}}
%PRESTO~\citep{2001PhDT.......123R, 2002AJ....124.1788R,2011ascl.soft07017R}, HEIMDALL~\citep[][\url{ https://sourceforge.net/projects/heimdall-astro/}]{heimdall}},PSRCHIVE~\citep[][\url{https://psrchive.sourceforge.net/ }]{psrchive}

%\software{\sc }
%\end{acknowledgments}

\bibliographystyle{aasjournal}
\bibliography{refs}
\appendix \nobreak
\renewcommand{\thefigure}{A\arabic{figure}}
\renewcommand{\thetable}{A\arabic{table}}
\setcounter{figure}{0}
\setcounter{table}{0}
\begin{figure*}[h]
    \centering
    \includegraphics[width=\linewidth]{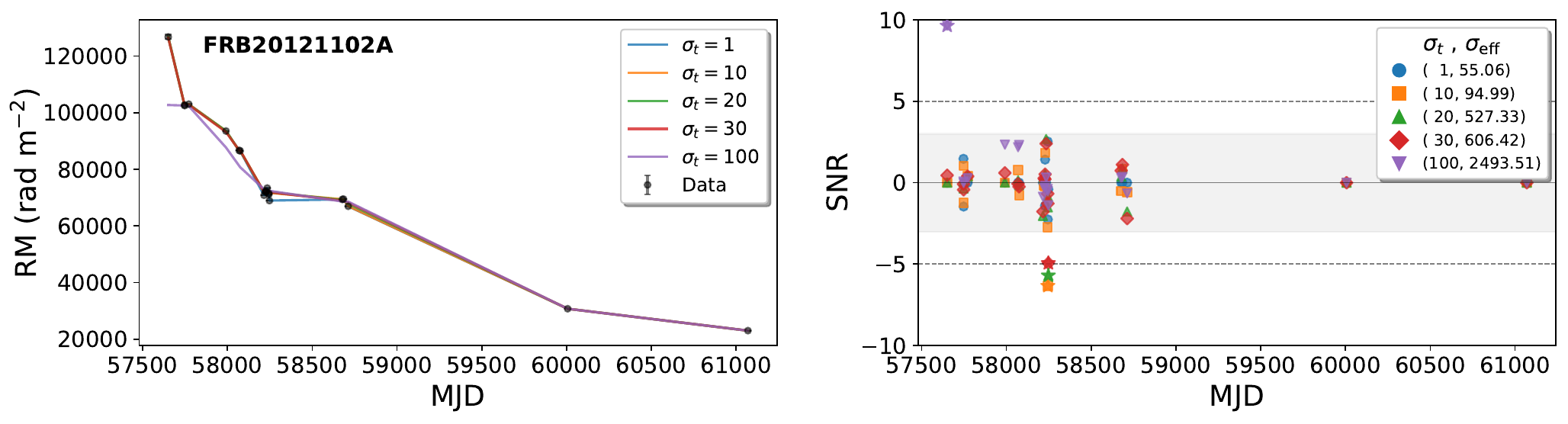}
    \includegraphics[width=\linewidth]{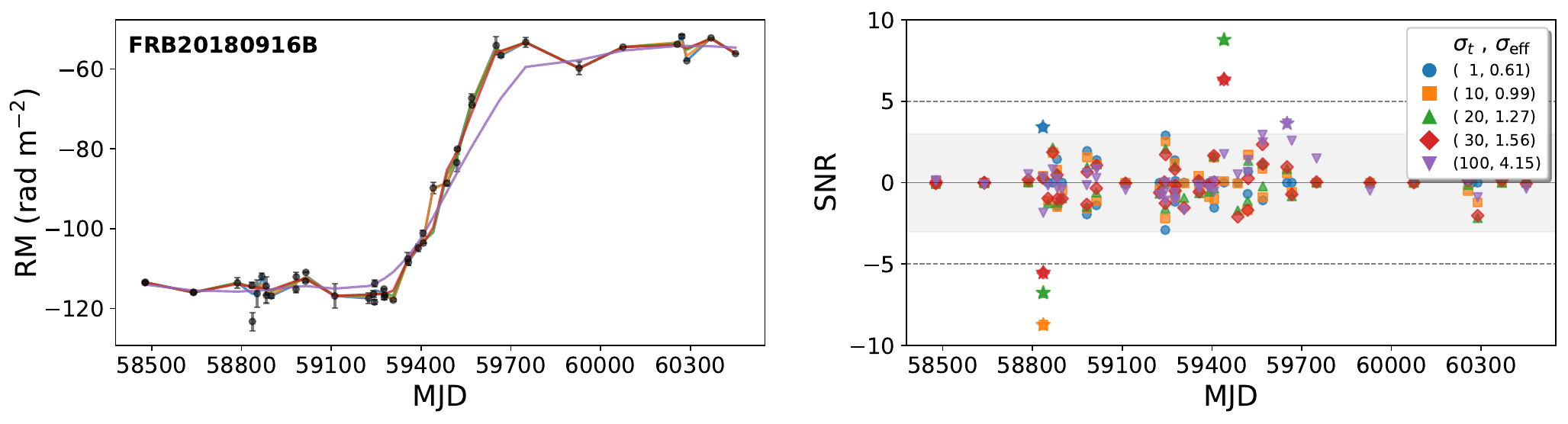}
    \includegraphics[width=\linewidth]{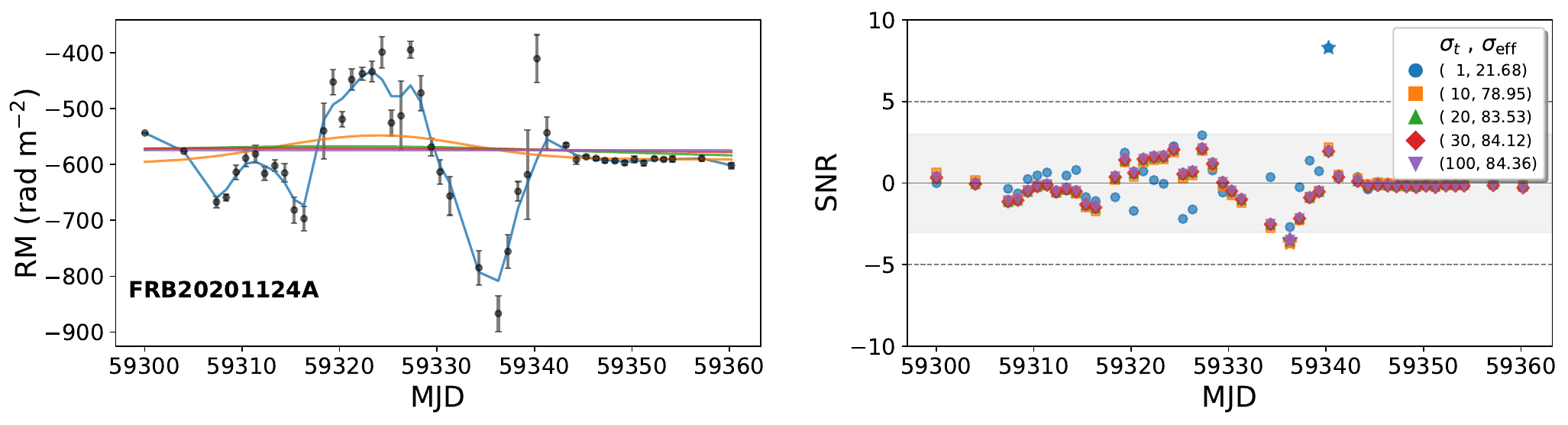}
    \includegraphics[width=\linewidth]{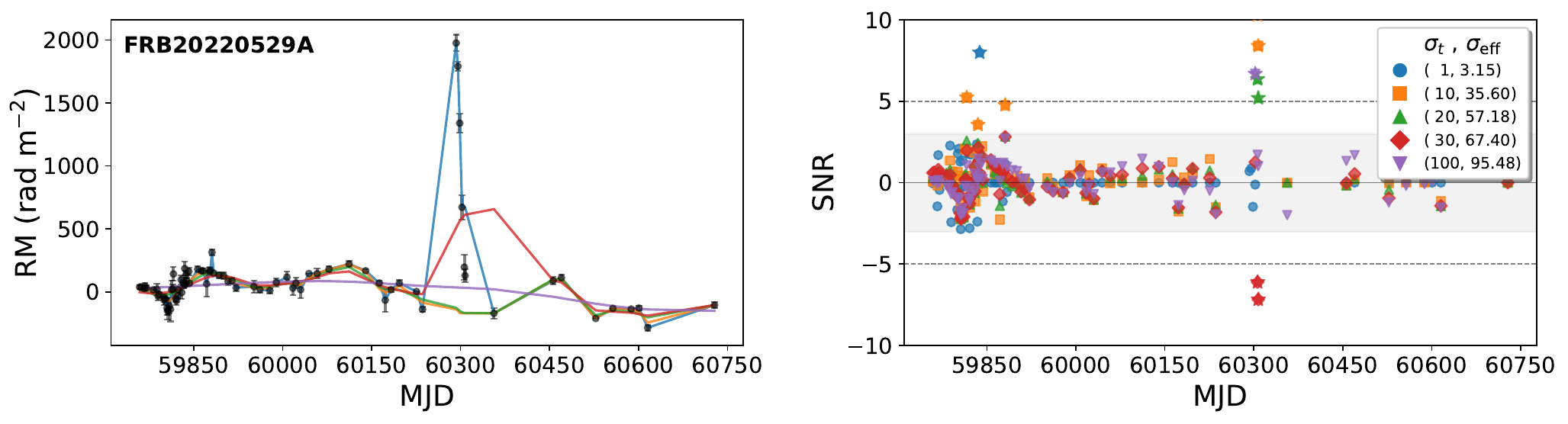}
    \caption{Effect of different $\sigma_t$ in Gaussian smoothing with outlier rejection. Left: Estimated baselines for different Gaussian smoothing width $\sigma_t$. Right: Significance of different $\sigma_t$, with the corresponding colors in the left panels.}
    \label{fig:sigmat_comparison}
\end{figure*}

\begin{figure*}[h]
    \centering
    \includegraphics[width=\linewidth]{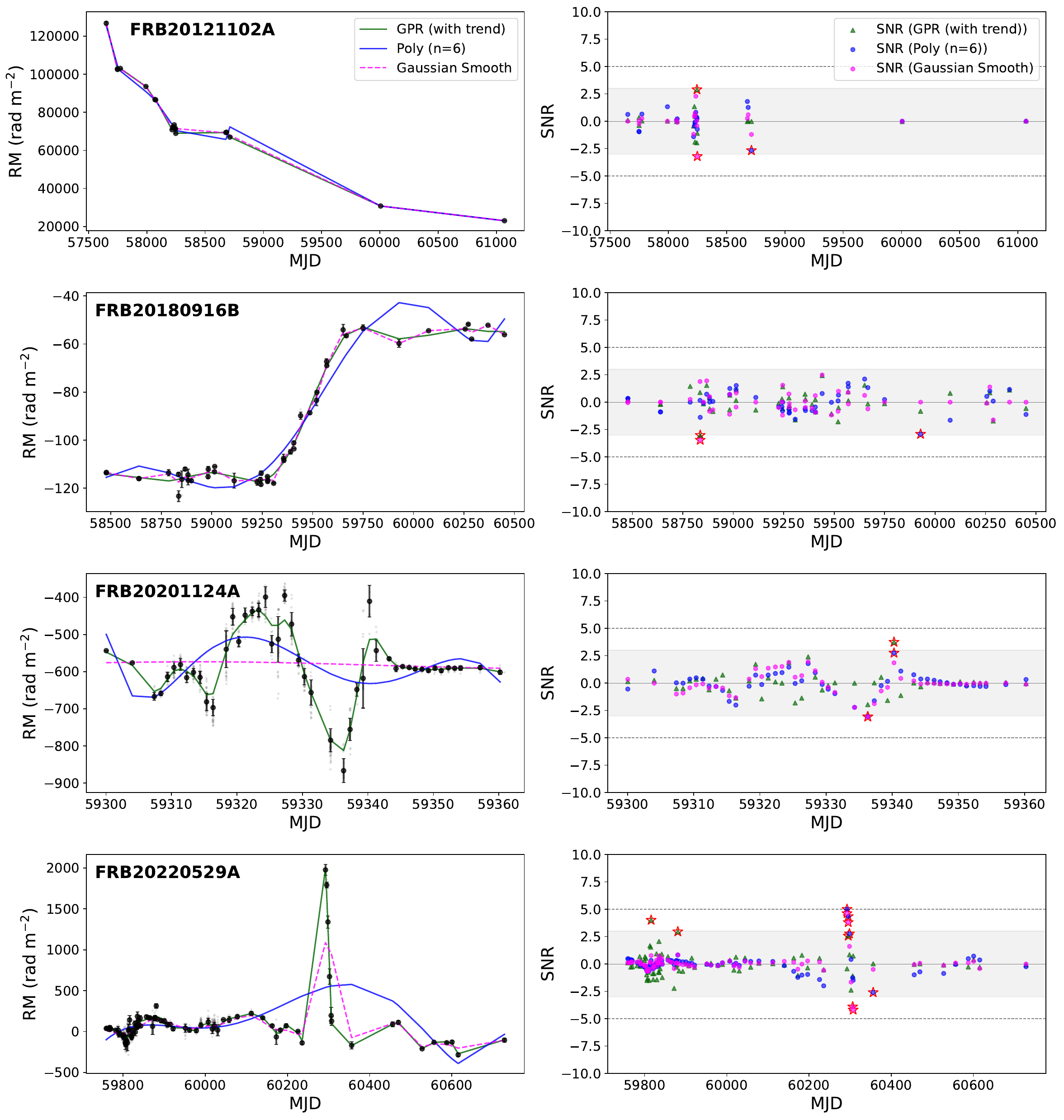}
    \caption{Effect of different baseline methods: Gaussian process regression with detrending (GPR), polynomial fitting (Poly, order 6), and Gaussian smoothing without iteration (width $\sigma_t$ = 20 days).}
    \label{fig:method_comparison}
\end{figure*}

\end{document}

%% file: sample.tex
\begin{table*}[htbp]
\centering
\caption{Summary of repeating FRB RM observations used in this work.}
\label{tab:rm_summary}
\begin{tabular}{lcccccccccc}
\hline
FRB & Telescope & $N_{\rm RM}$ & $N_{\rm epoch}$ & MJD range & \multicolumn{2}{c}{raw RM} & \multicolumn{2}{c}{daily binned RM} & References \\
 & & & & & median & $\sigma_{\rm RM}$ & median & $\sigma_{\rm RM}$ & \\
\hline
20121102A & Arecibo, Effelsberg & 37 & 20 & 57651--61070 & 72248 & 21622 & 72076 & 22841 & 1, 4, 10, 13, 15 \\
 & VLA, FAST, EVN & & & & & & & & \\
20180916B & CHIME & 54 & 46 & 58477--60451 & -112 & 24 & -112 & 25 & 8, 11 \\
20190303A & FAST, CHIME & 26 & 20 & 58666--60283 & -564 & 307 & -531 & 312 & 3, 9, 11 \\
20180301A & Parkes, FAST & 20 & 13 & 58763--59757 & -89 & 326 & -122 & 253 & 5, 7 \\
20190208A & FAST, CHIME & 13 & 10 & 58873--60010 & 26 & 7 & 26 & 8 & 3, 9, 11 \\
20190417A & CHIME, FAST & 39 & 15 & 59056--59805 & 4740 & 365 & 4754 & 306 & 3, 9 \\
20190520B & GBT, Parkes & 13 & 8 & 59107--59589 & 3888 & 13545 & 2750 & 12191 & 2 \\
20201124A & FAST, CHIME & 1106 & 47 & 59300--59361 & -593 & 106 & -591 & 94 & 11, 14 \\
20220529A & FAST & 666 & 81 & 59759--60728 & 15 & 242 & 46 & 343 & 6 \\
20220912A & FAST & 913 & 17 & 59880--59935 & -0 & 6 & 0 & 1 & 16 \\
\hline
\end{tabular}
\tablecomments{References:
$1.$ \cite{Plavin2022};
$2.$ \cite{Anna-Thomas2023};
$3.$ \cite{Fengy2025};
$4.$ \cite{Hilmarsson2021};
$5.$ \cite{kumar2023};
$6.$ \cite{LiY2026};
$7.$ \cite{Luor2020};
$8.$ \cite{Mckinven2023};
$9.$ \cite{Mckinven2023p};
$10.$ \cite{Michilli2018};
$11.$ \cite{Ng2025};
$12.$ \cite{Nimmo2022};
$13.$ \cite{Wang2025};
$14.$ \cite{XuH2022};
$15.$ \cite{ZhangJS2026};
$16.$ \cite{Zhangyk2023};
}
\end{table*}

%% file: flare_candidates.tex
\begin{table*}[htbp]\centering
\caption{RM flare candidates.}
\label{tab:flare}
\begin{tabular}{l|cccc|ccc|c}\hline
FRB & \multicolumn{4}{c|}{Initial candidate} & \multicolumn{4}{c}{Gaussian fit} \\\hline
 & MJD & RM & Baseline & SNR & MJD & A & $\sigma$ & comments \\ \hline    & days & rad m$^{-2}$ & rad m$^{-2}$ & day & rad m$^{-2}$ & day &  \\ \hline
FRB20121102A & 58247.8 & $68937.0\pm72.0$ & $71945.2$ & $-5.7$ & $58231.9\pm1.3$ & $6252 \pm 37482$ & $20 \pm 70$ & unstable baseline\\
FRB20180916B & 58836.2 & $-123.3\pm2.3$ & $-114.7$ & $-6.8$ & $58836.1\pm0.1$ & $-10\pm3$ & $0.5\pm0.1$ & 1 epoch\\
 & 59440.5 & $-89.8\pm1.5$ & $-101.0$ & $8.8$ & $59440.5^*$ & $8$ & $4$ & 1 epoch \\
FRB20201124A & 59336.3 & $-866.7\pm32.1$ & $-571.6$ & $-3.5$ & $59335.2\pm2.8$ & $-297\pm355$ & $2.0\pm2.5$ & alternative model\\
FRB20220529A & 60292.6 & $1976.6\pm66.5$ & $-127.1$ & $36.8$  & $60292.5\pm 0.1$ & $2046\pm19$ & $6.9\pm 0.1$ & \\
 & 60295.6 & $1791.2\pm35.5$ & $-138.8$ & $33.8$ & & & \\
 & 60298.5 & $1340.0\pm74.0$ & $-151.4$ & $26.1$ & & & \\
 & 60302.4 & $672.0\pm96.0$ & $-162.7$ & $14.6$ & & & \\
 & 60306.4 & $196.0\pm99.0$ & $-167.7$ & $6.4$ & & & \\
 & 60307.6 & $129.0\pm53.0$ & $-168.4$ & $5.2$ & & & \\
\hline\end{tabular}
\noindent {\small $^*$ The parameters are not constrained.}
\end{table*}

%% file: refs.bib
@ARTICLE{kumar2023,
       author = {{Kumar}, P. and {Luo}, R. and {Price}, D.~C. and {Shannon}, R.~M. and {Deller}, A.~T. and {Bhandari}, S. and {Feng}, Y. and {Flynn}, C. and {Jiang}, J.~C. and {Uttarkar}, P.~A. and {Wang}, S.~Q. and {Zhang}, S.~B.},
        title = "{Spectropolarimetric variability in the repeating fast radio burst source FRB 20180301A}",
      journal = {\mnras},
         year = 2023,
        month = dec,
       volume = {526},
       number = {3},
        pages = {3652-3672},
          doi = {10.1093/mnras/stad2969},
archivePrefix = {arXiv},
       eprint = {2304.01763},
 primaryClass = {astro-ph.HE},
       adsurl = {https://ui.adsabs.harvard.edu/abs/2023MNRAS.526.3652K}
}

@ARTICLE{Fengy2025,
       author = {{Feng}, Yi and {Zhang}, Yong-Kun and {Xie}, Jintao and {Yang}, Yuan-Pei and {Qu}, Yuanhong and {Zhou}, Dengke and {Li}, Di and {Zhang}, Bing and {Zhu}, Weiwei and {Lu}, Wenbin and {Xu}, Jiaying and {Miao}, Chenchen and {Tian}, Shiyan and {Wang}, Pei and {Yao}, Ju-Mei and {Niu}, Chen-Hui and {Niu}, Jiarui and {Xu}, Heng and {Jiang}, Jinchen and {Zhou}, Dejiang and {Liu}, Zenan and {Tsai}, Chao-Wei and {Dai}, Zigao and {Wu}, Xuefeng and {Wang}, Fayin and {Han}, Jinlin and {Lee}, Kejia and {Xu}, Renxin and {Huang}, Yongfeng and {Zou}, Yuanchuan and {Cao}, Jinhuang and {Chen}, Xianglei and {Fang}, Jianhua and {Li}, Dongzi and {Li}, Ye and {Lu}, Wanjin and {Luo}, Jiawei and {Luo}, Jintao and {Luo}, Rui and {Lyu}, Fen and {Wang}, Bojun and {Wang}, Weiyang and {Wu}, Qin and {Xue}, Mengyao and {Xiao}, Di and {Yu}, Wenfei and {Yuan}, Jianping and {Zhang}, Chunfeng and {Zhang}, Junshuo and {Zhang}, Lei and {Zhang}, Songbo and {Zhao}, Rushuang and {Zhu}, Yuhao},
        title = "{Multi-year polarimetric monitoring of four CHIME-discovered repeating fast radio bursts with FAST}",
      journal = {Science China Physics, Mechanics, and Astronomy},
         year = 2025,
        month = aug,
       volume = {68},
       number = {8},
          eid = {289511},
        pages = {289511},
          doi = {10.1007/s11433-024-2668-5},
archivePrefix = {arXiv},
       eprint = {2507.02355},
 primaryClass = {astro-ph.HE},
       adsurl = {https://ui.adsabs.harvard.edu/abs/2025SCPMA..6889511F}
}

@ARTICLE{ZhangJS2026,
       author = {{Zhang}, Junshuo and {Zhang}, Yongkun and {Feng}, Yi and {Liang}, Chengwei and {Cao}, Shuo and {Zhang}, Jiaheng and {Wang}, Pei and {Wang}, Tiancong and {Zhou}, Dejiang and {Niu}, Jiarui and {Di Li} and {Zhu}, Weiwei and {Zhang}, Bing and {Han}, Jinlin and {Zou}, Yuanchuan and {Luo}, Rui and {Niu}, Chenhui and {Li}, Ye and {Yu}, Wenfei and {Zhang}, Songbo and {Lu}, Wanjin and {Cao}, Jinhuang and {Wang}, Yidan and {Zhou}, Dengke and {Zhu}, Yuhao},
        title = "{Detection of high activity from repeating FRB 20121102A at L-band with FAST}",
      journal = {The Astronomer's Telegram},
         year = 2026,
        month = feb,
       volume = {17642},
        pages = {1},
       adsurl = {https://ui.adsabs.harvard.edu/abs/2026ATel17642....1Z}
}

@ARTICLE{Wang2025,
       author = {{Wang}, P. and {Zhang}, J.~S. and {Yang}, Y.~P. and {Zhou}, D.~K. and {Zhang}, Y.~K. and {Feng}, Y. and {Zhao}, Z.~Y. and {Fang}, J.~H. and {Li}, D. and {Zhu}, W.~W. and et al.},
        title = "{Decadal evolution of a repeating fast radio burst source}",
      journal = {arXiv e-prints},
         year = 2025,
        month = jul,
          eid = {arXiv:2507.15790},
        pages = {arXiv:2507.15790},
          doi = {10.48550/arXiv.2507.15790},
archivePrefix = {arXiv},
       eprint = {2507.15790},
 primaryClass = {astro-ph.HE},
       adsurl = {https://ui.adsabs.harvard.edu/abs/2025arXiv250715790W}
}

@ARTICLE{Plavin2022,
       author = {{Plavin}, A. and {Paragi}, Z. and {Marcote}, B. and {Keimpema}, A. and {Hessels}, J.~W.~T. and {Nimmo}, K. and {Vedantham}, H.~K. and {Spitler}, L.~G.},
        title = "{FRB 121102: Drastic changes in the burst polarization contrasts with the stability of the persistent emission}",
      journal = {\mnras},
         year = 2022,
        month = apr,
       volume = {511},
       number = {4},
        pages = {6033-6041},
          doi = {10.1093/mnras/stac500},
archivePrefix = {arXiv},
       eprint = {2202.10519},
 primaryClass = {astro-ph.HE},
       adsurl = {https://ui.adsabs.harvard.edu/abs/2022MNRAS.511.6033P}
}

@ARTICLE{Ng2025,
       author = {{Ng}, Cherry and {Pandhi}, Ayush and {Mckinven}, Ryan and {Curtin}, Alice P. and {Shin}, Kaitlyn and {Fonseca}, Emmanuel and {Gaensler}, B.~M. and {Jow}, Dylan L. and {Kaspi}, Victoria and {Li}, Dongzi and {Main}, Robert and {Masui}, Kiyoshi W. and {Michilli}, Daniele and {Nimmo}, Kenzie and {Pleunis}, Ziggy and {Scholz}, Paul and {Stairs}, Ingrid and {Bhardwaj}, Mohit and {Brar}, Charanjot and {Cassanelli}, Tomas and {Joseph}, Ronniy C. and {Pearlman}, Aaron B. and {Rafiei-Ravandi}, Masoud and {Smith}, Kendrick},
        title = "{Polarization Properties of 28 Repeating Fast Radio Burst Sources with CHIME/FRB}",
      journal = {\apj},
         year = 2025,
        month = apr,
       volume = {982},
       number = {2},
          eid = {154},
        pages = {154},
          doi = {10.3847/1538-4357/adb0bc},
archivePrefix = {arXiv},
       eprint = {2411.09045},
 primaryClass = {astro-ph.HE},
       adsurl = {https://ui.adsabs.harvard.edu/abs/2025ApJ...982..154N}
}

@ARTICLE{astropy:2018,
   author = {{Astropy Collaboration} and {Price-Whelan}, A.~M. and {Sip{\H o}cz}, B.~M. and
	{G{\"u}nther}, H.~M. and {Lim}, P.~L. and {Crawford}, S.~M. and
	{Conseil}, S. and {Shupe}, D.~L. and {Craig}, M.~W. and {Dencheva}, N. and
	{Ginsburg}, A. and {VanderPlas}, J.~T. and {Bradley}, L.~D. and
	{P{\'e}rez-Su{\'a}rez}, D. and {de Val-Borro}, M. and {Paper Contributors}, (. and
	{Aldcroft}, T.~L. and {Cruz}, K.~L. and {Robitaille}, T.~P. and
	{Tollerud}, E.~J. and {Coordination Committee}, (. and {Ardelean}, C. and
	{Babej}, T. and {Bach}, Y.~P. and {Bachetti}, M. and {Bakanov}, A.~V. and
	{Bamford}, S.~P. and {Barentsen}, G. and {Barmby}, P. and {Baumbach}, A. and
	{Berry}, K.~L. and {Biscani}, F. and {Boquien}, M. and {Bostroem}, K.~A. and
	{Bouma}, L.~G. and {Brammer}, G.~B. and {Bray}, E.~M. and {Breytenbach}, H. and
	{Buddelmeijer}, H. and {Burke}, D.~J. and {Calderone}, G. and
	{Cano Rodr{\'{\i}}guez}, J.~L. and {Cara}, M. and {Cardoso}, J.~V.~M. and
	{Cheedella}, S. and {Copin}, Y. and {Corrales}, L. and {Crichton}, D. and
	{D{\'A}vella}, D. and {Deil}, C. and {Depagne}, {\'E}. and
	{Dietrich}, J.~P. and {Donath}, A. and {Droettboom}, M. and
	{Earl}, N. and {Erben}, T. and {Fabbro}, S. and {Ferreira}, L.~A. and
	{Finethy}, T. and {Fox}, R.~T. and {Garrison}, L.~H. and {Gibbons}, S.~L.~J. and
	{Goldstein}, D.~A. and {Gommers}, R. and {Greco}, J.~P. and
	{Greenfield}, P. and {Groener}, A.~M. and {Grollier}, F. and
	{Hagen}, A. and {Hirst}, P. and {Homeier}, D. and {Horton}, A.~J. and
	{Hosseinzadeh}, G. and {Hu}, L. and {Hunkeler}, J.~S. and {Ivezi{\'c}}, {\v Z}. and
	{Jain}, A. and {Jenness}, T. and {Kanarek}, G. and {Kendrew}, S. and
	{Kern}, N.~S. and {Kerzendorf}, W.~E. and {Khvalko}, A. and
	{King}, J. and {Kirkby}, D. and {Kulkarni}, A.~M. and {Kumar}, A. and
	{Lee}, A. and {Lenz}, D. and {Littlefair}, S.~P. and {Ma}, Z. and
	{Macleod}, D.~M. and {Mastropietro}, M. and {McCully}, C. and
	{Montagnac}, S. and {Morris}, B.~M. and {Mueller}, M. and {Mumford}, S.~J. and
	{Muna}, D. and {Murphy}, N.~A. and {Nelson}, S. and {Nguyen}, G.~H. and
	{Ninan}, J.~P. and {N{\"o}the}, M. and {Ogaz}, S. and {Oh}, S. and
	{Parejko}, J.~K. and {Parley}, N. and {Pascual}, S. and {Patil}, R. and
	{Patil}, A.~A. and {Plunkett}, A.~L. and {Prochaska}, J.~X. and
	{Rastogi}, T. and {Reddy Janga}, V. and {Sabater}, J. and {Sakurikar}, P. and
	{Seifert}, M. and {Sherbert}, L.~E. and {Sherwood-Taylor}, H. and
	{Shih}, A.~Y. and {Sick}, J. and {Silbiger}, M.~T. and {Singanamalla}, S. and
	{Singer}, L.~P. and {Sladen}, P.~H. and {Sooley}, K.~A. and
	{Sornarajah}, S. and {Streicher}, O. and {Teuben}, P. and {Thomas}, S.~W. and
	{Tremblay}, G.~R. and {Turner}, J.~E.~H. and {Terr{\'o}n}, V. and
	{van Kerkwijk}, M.~H. and {de la Vega}, A. and {Watkins}, L.~L. and
	{Weaver}, B.~A. and {Whitmore}, J.~B. and {Woillez}, J. and
	{Zabalza}, V. and {Contributors}, (.},
    title = "{The Astropy Project: Building an Open-science Project and Status of the v2.0 Core Package}",
  journal = {\aj},
archivePrefix = "arXiv",
   eprint = {1801.02634},
 primaryClass = "astro-ph.IM",
     year = 2018,
    month = sep,
   volume = 156,
      eid = {123},
    pages = {123},
      doi = {10.3847/1538-3881/aabc4f},
   adsurl = {https://ui.adsabs.harvard.edu/abs/2018AJ....156..123T}
}

@ARTICLE{astropy:2013,
   author = {{Astropy Collaboration} and {Robitaille}, T.~P. and {Tollerud}, E.~J. and
    {Greenfield}, P. and {Droettboom}, M. and {Bray}, E. and {Aldcroft}, T. and
    {Davis}, M. and {Ginsburg}, A. and {Price-Whelan}, A.~M. and
    {Kerzendorf}, W.~E. and {Conley}, A. and {Crighton}, N. and
    {Barbary}, K. and {Muna}, D. and {Ferguson}, H. and {Grollier}, F. and
    {Parikh}, M.~M. and {Nair}, P.~H. and {Unther}, H.~M. and {Deil}, C. and
    {Woillez}, J. and {Conseil}, S. and {Kramer}, R. and {Turner}, J.~E.~H. and
    {Singer}, L. and {Fox}, R. and {Weaver}, B.~A. and {Zabalza}, V. and
    {Edwards}, Z.~I. and {Azalee Bostroem}, K. and {Burke}, D.~J. and
    {Casey}, A.~R. and {Crawford}, S.~M. and {Dencheva}, N. and
    {Ely}, J. and {Jenness}, T. and {Labrie}, K. and {Lian Lim}, P. and
    {Pierfederici}, F. and {Pontzen}, A. and {Ptak}, A. and {Refsdal}, B. and
    {Servillat}, M. and {Streicher}, O.},
    title = "{Astropy: A community Python package for astronomy}",
  journal = {\aap},
     year = 2013,
    month = oct,
   volume = 558,
      eid = {A33},
    pages = {A33},
      doi = {10.1051/0004-6361/201322068},
   adsurl = {https://ui.adsabs.harvard.edu/abs/2013A%26A...558A..33A}
}

@ARTICLE{Luor2020,
       author = {{Luo}, R. and {Wang}, B.~J. and {Men}, Y.~P. and {Zhang}, C.~F. and {Jiang}, J.~C. and {Xu}, H. and {Wang}, W.~Y. and {Lee}, K.~J. and {Han}, J.~L. and {Zhang}, B. and {Caballero}, R.~N. and {Chen}, M.~Z. and {Chen}, X.~L. and {Gan}, H.~Q. and {Guo}, Y.~J. and {Hao}, L.~F. and {Huang}, Y.~X. and {Jiang}, P. and {Li}, H. and {Li}, J. and {Li}, Z.~X. and {Luo}, J.~T. and {Pan}, J. and {Pei}, X. and {Qian}, L. and {Sun}, J.~H. and {Wang}, M. and {Wang}, N. and {Wen}, Z.~G. and {Xu}, R.~X. and {Xu}, Y.~H. and {Yan}, J. and {Yan}, W.~M. and {Yu}, D.~J. and {Yuan}, J.~P. and {Zhang}, S.~B. and {Zhu}, Y.},
        title = "{Diverse polarization angle swings from a repeating fast radio burst source}",
      journal = {\nat},
         year = 2020,
        month = oct,
       volume = {586},
       number = {7831},
        pages = {693-696},
          doi = {10.1038/s41586-020-2827-2},
archivePrefix = {arXiv},
       eprint = {2011.00171},
 primaryClass = {astro-ph.HE},
       adsurl = {https://ui.adsabs.harvard.edu/abs/2020Natur.586..693L}
}

@ARTICLE{Zhoudj2025,
       author = {{Zhou}, DeJiang and {Han}, J.~L. and {Zhang}, Bing and {Zhu}, WeiWei and {Wang}, Wei-yang and {Yang}, Yuan-Pei and {Qu}, Yuanhong and {Zhang}, Yong-Kun and {Yan}, Yi and {Jing}, Wei-Cong and {Cao}, Shuo and {Xie}, Jintao and {Yang}, Xuan and {Tian}, Shiyan and {Li}, Ye and {Li}, Dongzi and {Niu}, Jia-Rui and {Wu}, Zi-Wei and {Wu}, Qin and {Feng}, Yi and {Wang}, Fayin and {Wang}, Pei},
        title = "{Bright bursts with sub-millisecond structures of FRB 20230607A in a highly magnetized environment}",
      journal = {arXiv e-prints},
         year = 2025,
        month = apr,
          eid = {arXiv:2504.11173},
        pages = {arXiv:2504.11173},
          doi = {10.48550/arXiv.2504.11173},
archivePrefix = {arXiv},
       eprint = {2504.11173},
 primaryClass = {astro-ph.HE},
       adsurl = {https://ui.adsabs.harvard.edu/abs/2025arXiv250411173Z}
}

@article{Thornton2013,
   title={A Population of Fast Radio Bursts at Cosmological Distances},
   volume={341},
   ISSN={1095-9203},
   url={http://dx.doi.org/10.1126/science.1236789},
   DOI={10.1126/science.1236789},
   number={6141},
   journal={Science},
   publisher={American Association for the Advancement of Science (AAAS)},
   author={Thornton, D. and Stappers, B. and Bailes, M. and Barsdell, B. and Bates, S. and Bhat, N. D. R. and Burgay, M. and Burke-Spolaor, S. and Champion, D. J. and Coster, P. and D’Amico, N. and Jameson, A. and Johnston, S. and Keith, M. and Kramer, M. and Levin, L. and Milia, S. and Ng, C. and Possenti, A. and van Straten, W.},
   year={2013},
   month=jul, pages={53–56} }

@misc{Zhangyk2023,
      title={FAST Observations of FRB 20220912A: Burst Properties and Polarization Characteristics}, 
      author={Yong-Kun Zhang and Di Li and Bing Zhang and Shuo Cao and Yi Feng and Wei-Yang Wang and Yuan-Hong Qu and Jia-Rui Niu and Wei-Wei Zhu and Jin-Lin Han and Peng Jiang and Ke-Jia Lee and Dong-Zi Li and Rui Luo and Chen-Hui Niu and Chao-Wei Tsai and Pei Wang and Fa-Yin Wang and Zi-Wei Wu and Heng Xu and Yuan-Pei Yang and Jun-Shuo Zhang and De-Jiang Zhou and Yu-Hao Zhu},
      year={2023},
      eprint={2304.14665},
      archivePrefix={arXiv},
      primaryClass={astro-ph.HE},
      url={https://arxiv.org/abs/2304.14665}, 
}

@article{Zhoudj2022,
   title={FAST Observations of an Extremely Active Episode of FRB 20201124A: I. Burst Morphology},
   volume={22},
   ISSN={1674-4527},
   url={http://dx.doi.org/10.1088/1674-4527/ac98f8},
   DOI={10.1088/1674-4527/ac98f8},
   number={12},
   journal={Research in Astronomy and Astrophysics},
   publisher={IOP Publishing},
   author={Zhou, D. J. and Han, J. L. and Zhang, B. and Lee, K. J. and Zhu, W. W. and Li, D. and Jing, W. C. and Wang, W. -Y. and Zhang, Y. K. and Jiang, J. C. and Niu, J. R. and Luo, R. and Xu, H. and Zhang, C. F. and Wang, B. J. and Xu, J. W. and Wang, P. and Yang, Z. L. and Feng, Y.},
   year={2022},
   month=nov, pages={124001} }

@misc{XuJW2025,
      title={Periodic variation of magnetoionic environment of a fast radio burst source}, 
      author={Jiangwei Xu and Heng Xu and Yanjun Guo and Jinchen Jiang and Bojun Wang and Zihan Xue and Yunpeng Men and Kejia Lee and Bing Zhang and Weiwei Zhu and Jinlin Han},
      year={2025},
      eprint={2505.06006},
      archivePrefix={arXiv},
      primaryClass={astro-ph.HE},
      url={https://arxiv.org/abs/2505.06006}, 
}

@ARTICLE{Anna-Thomas2023,
       author = {{Anna-Thomas}, Reshma and {Connor}, Liam and {Dai}, Shi and {Feng}, Yi and {Burke-Spolaor}, Sarah and {Beniamini}, Paz and {Yang}, Yuan-Pei and {Zhang}, Yong-Kun and {Aggarwal}, Kshitij and {Law}, Casey J. and {Li}, Di and {Niu}, Chenhui and {Chatterjee}, Shami and {Cruces}, Marilyn and {Duan}, Ran and {Filipovic}, Miroslav D. and {Hobbs}, George and {Lynch}, Ryan S. and {Miao}, Chenchen and {Niu}, Jiarui and {Ocker}, Stella K. and {Tsai}, Chao-Wei and {Wang}, Pei and {Xue}, Mengyao and {Yao}, Ju-Mei and {Yu}, Wenfei and {Zhang}, Bing and {Zhang}, Lei and {Zhu}, Shiqiang and {Zhu}, Weiwei},
        title = "{Magnetic field reversal in the turbulent environment around a repeating fast radio burst}",
      journal = {Science},
         year = 2023,
        month = may,
       volume = {380},
       number = {6645},
        pages = {599-603},
          doi = {10.1126/science.abo6526},
archivePrefix = {arXiv},
       eprint = {2202.11112},
 primaryClass = {astro-ph.HE},
       adsurl = {https://ui.adsabs.harvard.edu/abs/2023Sci...380..599A}
}

@ARTICLE{wangfy2022,
       author = {{Wang}, F.~Y. and {Zhang}, G.~Q. and {Dai}, Z.~G. and {Cheng}, K.~S.},
        title = "{Repeating fast radio burst 20201124A originates from a magnetar/Be star binary}",
      journal = {Nature Communications},
         year = 2022,
        month = sep,
       volume = {13},
          eid = {4382},
        pages = {4382},
          doi = {10.1038/s41467-022-31923-y},
archivePrefix = {arXiv},
       eprint = {2204.08124},
 primaryClass = {astro-ph.HE},
       adsurl = {https://ui.adsabs.harvard.edu/abs/2022NatCo..13.4382W}
}

@article{chime180916,
   title={Periodic activity from a fast radio burst source},
   volume={582},
   ISSN={1476-4687},
   url={http://dx.doi.org/10.1038/s41586-020-2398-2},
   DOI={10.1038/s41586-020-2398-2},
   number={7812},
   journal={Nature},
   publisher={Springer Science and Business Media LLC},
   author={Amiri, M. and Andersen, B. C. and Bandura, K. M. and Bhardwaj, M. and Boyle, P. J. and Brar, C. and Chawla, P. and Chen, T. and Cliche, J. F. and Cubranic, D. and Deng, M. and Denman, N. T. and Dobbs, M. and Dong, F. Q. and Fandino, M. and Fonseca, E. and Gaensler, B. M. and Giri, U. and Good, D. C. and Halpern, M. and Hessels, J. W. T. and Hill, A. S. and Höfer, C. and Josephy, A. and Kania, J. W. and Karuppusamy, R. and Kaspi, V. M. and Keimpema, A. and Kirsten, F. and Landecker, T. L. and Lang, D. A. and Leung, C. and Li, D. Z. and Lin, H.-H. and Marcote, B. and Masui, K. W. and Mckinven, R. and Mena-Parra, J. and Merryfield, M. and Michilli, D. and Milutinovic, N. and Mirhosseini, A. and Naidu, A. and Newburgh, L. B. and Ng, C. and Nimmo, K. and Paragi, Z. and Patel, C. and Pen, U.-L. and Pinsonneault-Marotte, T. and Pleunis, Z. and Rafiei-Ravandi, M. and Rahman, M. and Ransom, S. M. and Renard, A. and Sanghavi, P. and Scholz, P. and Shaw, J. R. and Shin, K. and Siegel, S. R. and Singh, S. and Smegal, R. J. and Smith, K. M. and Stairs, I. H. and Tendulkar, S. P. and Tretyakov, I. and Vanderlinde, K. and Wang, H. and Wang, X. and Wulf, D. and Yadav, P. and Zwaniga, A. V.},
   year={2020},
   month=jun, pages={351–355} }

@ARTICLE{LiY2026,
       author = {{Li}, Y. and {Zhang}, S.~B. and {Yang}, Y.~P. and {Tsai}, C.~W. and {Yang}, X. and {Law}, C.~J. and {Anna-Thomas}, R. and {Chen}, X.~L. and {Lee}, K.~J. and {Tang}, Z.~F. and {Xiao}, D. and {Xu}, H. and {Yang}, X.~L. and {Chen}, G. and {Feng}, Y. and {Li}, D.~Z. and {Mckinven}, R. and {Niu}, J.~R. and {Shin}, K. and {Wang}, B.~J. and {Zhang}, C.~F. and {Zhang}, Y.~K. and {Zhou}, D.~J. and {Zhu}, Y.~H. and {Dai}, Z.~G. and {Chang}, C.~M. and {Geng}, J.~J. and {Han}, J.~L. and {Hu}, L. and {Li}, D. and {Luo}, R. and {Niu}, C.~H. and {Shi}, D.~D. and {Sun}, T.~R. and {Wu}, X.~F. and {Zhu}, W.~W. and {Jiang}, P. and {Zhang}, B.},
        title = "{A sudden change and recovery in the magnetic environment around a repeating fast radio burst}",
      journal = {Science},
         year = 2026,
        month = jan,
       volume = {391},
       number = {6782},
        pages = {280-284},
          doi = {10.1126/science.adq3225},
archivePrefix = {arXiv},
       eprint = {2503.04727},
 primaryClass = {astro-ph.HE},
       adsurl = {https://ui.adsabs.harvard.edu/abs/2026Sci...391..280L}
}

@article{Mckinven2023,
   title={A Large-scale Magneto-ionic Fluctuation in the Local Environment of Periodic Fast Radio Burst Source FRB 20180916B},
   volume={950},
   ISSN={1538-4357},
   url={http://dx.doi.org/10.3847/1538-4357/acc65f},
   DOI={10.3847/1538-4357/acc65f},
   number={1},
   journal={The Astrophysical Journal},
   publisher={American Astronomical Society},
   author={Mckinven, R. and Gaensler, B. M. and Michilli, D. and Masui, K. and Kaspi, V. M. and Bhardwaj, M. and Cassanelli, T. and Chawla, P. and Dong, F. (Adam) and Fonseca, E. and Leung, C. and Li, D. Z. and Ng, C. and Patel, C. and Petroff, E. and Pearlman, A. B. and Pleunis, Z. and Rafiei-Ravandi, M. and Rahman, M. and Sand, K. R. and Shin, K. and Scholz, P. and Stairs, I. H. and Smith, K. and Su, J. and Tendulkar, S.},
   year={2023},
   month=jun, pages={12} }

@article{Michilli2018,
   title={An extreme magneto-ionic environment associated with the fast radio burst source FRB 121102},
   volume={553},
   ISSN={1476-4687},
   url={http://dx.doi.org/10.1038/nature25149},
   DOI={10.1038/nature25149},
   number={7687},
   journal={Nature},
   publisher={Springer Science and Business Media LLC},
   author={Michilli, D. and Seymour, A. and Hessels, J. W. T. and Spitler, L. G. and Gajjar, V. and Archibald, A. M. and Bower, G. C. and Chatterjee, S. and Cordes, J. M. and Gourdji, K. and Heald, G. H. and Kaspi, V. M. and Law, C. J. and Sobey, C. and Adams, E. A. K. and Bassa, C. G. and Bogdanov, S. and Brinkman, C. and Demorest, P. and Fernandez, F. and Hellbourg, G. and Lazio, T. J. W. and Lynch, R. S. and Maddox, N. and Marcote, B. and McLaughlin, M. A. and Paragi, Z. and Ransom, S. M. and Scholz, P. and Siemion, A. P. V. and Tendulkar, S. P. and Van Rooy, P. and Wharton, R. S. and Whitlow, D.},
   year={2018},
   month=jan, pages={182–185} }

@article{Huber1964,
  author = {Huber, Peter J.},
  title = {Robust Estimation of a Location Parameter},
  journal = {The Annals of Mathematical Statistics},
  volume = {35},
  number = {1},
  pages = {73--101},
  year = {1964},
  doi = {10.1214/aoms/1177703732},
  url = {https://projecteuclid.org/journals/annals-of-mathematical-statistics/volume-35/issue-1}
}

@ARTICLE{FengY2022,
       author = {{Feng}, Yi and {Li}, Di and {Yang}, Yuan-Pei and {Zhang}, Yongkun and {Zhu}, Weiwei and {Zhang}, Bing and {Lu}, Wenbin and {Wang}, Pei and {Dai}, Shi and {Lynch}, Ryan S. and {Yao}, Jumei and {Jiang}, Jinchen and {Niu}, Jiarui and {Zhou}, Dejiang and {Xu}, Heng and {Miao}, Chenchen and {Niu}, Chenhui and {Meng}, Lingqi and {Qian}, Lei and {Tsai}, Chao-Wei and {Wang}, Bojun and {Xue}, Mengyao and {Yue}, Youling and {Yuan}, Mao and {Zhang}, Songbo and {Zhang}, Lei},
        title = "{Frequency-dependent polarization of repeating fast radio bursts{\textemdash}implications for their origin}",
      journal = {Science},
         year = 2022,
        month = mar,
       volume = {375},
       number = {6586},
        pages = {1266-1270},
          doi = {10.1126/science.abl7759},
archivePrefix = {arXiv},
       eprint = {2202.09601},
 primaryClass = {astro-ph.HE},
       adsurl = {https://ui.adsabs.harvard.edu/abs/2022Sci...375.1266F}
}

@ARTICLE{Johnston1996,
       author = {{Johnston}, Simon and {Manchester}, R.~N. and {Lyne}, A.~G. and {D'Amico}, N. and {Bailes}, M. and {Gaensler}, B.~M. and {Nicastro}, L.},
        title = "{Radio observations of PSR B1259-63 around periastron}",
      journal = {\mnras},
         year = 1996,
        month = apr,
       volume = {279},
       number = {3},
        pages = {1026-1036},
          doi = {10.1093/mnras/279.3.1026},
       adsurl = {https://ui.adsabs.harvard.edu/abs/1996MNRAS.279.1026J}
}

@ARTICLE{Kooi2017,
       author = {{Kooi}, Jason E. and {Fischer}, Patrick D. and {Buffo}, Jacob J. and {Spangler}, Steven R.},
        title = "{VLA Measurements of Faraday Rotation through Coronal Mass Ejections}",
      journal = {\solphys},
         year = 2017,
        month = apr,
       volume = {292},
       number = {4},
          eid = {56},
        pages = {56},
          doi = {10.1007/s11207-017-1074-7},
archivePrefix = {arXiv},
       eprint = {1611.01445},
 primaryClass = {astro-ph.SR},
       adsurl = {https://ui.adsabs.harvard.edu/abs/2017SoPh..292...56K}
}

@article{Cleveland1979,
  author = {Cleveland, William S.},
  title = {Robust Locally Weighted Regression and Smoothing Scatterplots},
  journal = {Journal of the American Statistical Association},
  volume = {74},
  number = {368},
  pages = {829--836},
  year = {1979},
  doi = {10.1080/01621459.1979.10481038},
  url = {https://www.tandfonline.com/doi/abs/10.1080/01621459.1979.10481038}
}

@article{Hilmarsson2021,
   title={Rotation Measure Evolution of the Repeating Fast Radio Burst Source FRB 121102},
   volume={908},
   ISSN={2041-8213},
   url={http://dx.doi.org/10.3847/2041-8213/abdec0},
   DOI={10.3847/2041-8213/abdec0},
   number={1},
   journal={The Astrophysical Journal Letters},
   publisher={American Astronomical Society},
   author={Hilmarsson, G. H. and Michilli, D. and Spitler, L. G. and Wharton, R. S. and Demorest, P. and Desvignes, G. and Gourdji, K. and Hackstein, S. and Hessels, J. W. T. and Nimmo, K. and Seymour, A. D. and Kramer, M. and Mckinven, R.},
   year={2021},
   month=feb, pages={L10} }

@article{XuH2022,
   title={A fast radio burst source at a complex magnetized site in a barred galaxy},
   volume={609},
   ISSN={1476-4687},
   url={http://dx.doi.org/10.1038/s41586-022-05071-8},
   DOI={10.1038/s41586-022-05071-8},
   number={7928},
   journal={Nature},
   publisher={Springer Science and Business Media LLC},
   author={Xu, H. and Niu, J. R. and Chen, P. and Lee, K. J. and Zhu, W. W. and Dong, S. and Zhang, B. and Jiang, J. C. and Wang, B. J. and Xu, J. W. and Zhang, C. F. and Fu, H. and Filippenko, A. V. and Peng, E. W. and Zhou, D. J. and Zhang, Y. K. and Wang, P. and Feng, Y. and Li, Y. and Brink, T. G. and Li, D. Z. and Lu, W. and Yang, Y. P. and Caballero, R. N. and Cai, C. and Chen, M. Z. and Dai, Z. G. and Djorgovski, S. G. and Esamdin, A. and Gan, H. Q. and Guhathakurta, P. and Han, J. L. and Hao, L. F. and Huang, Y. X. and Jiang, P. and Li, C. K. and Li, D. and Li, H. and Li, X. Q. and Li, Z. X. and Liu, Z. Y. and Luo, R. and Men, Y. P. and Niu, C. H. and Peng, W. X. and Qian, L. and Song, L. M. and Stern, D. and Stockton, A. and Sun, J. H. and Wang, F. Y. and Wang, M. and Wang, N. and Wang, W. Y. and Wu, X. F. and Xiao, S. and Xiong, S. L. and Xu, Y. H. and Xu, R. X. and Yang, J. and Yang, X. and Yao, R. and Yi, Q. B. and Yue, Y. L. and Yu, D. J. and Yu, W. F. and Yuan, J. P. and Zhang, B. B. and Zhang, S. B. and Zhang, S. N. and Zhao, Y. and Zheng, W. K. and Zhu, Y. and Zou, J. H.},
   year={2022},
   month=sep, pages={685–688} }

@ARTICLE{nimmo2022,
      author = {{Nimmo}, K. and {Hewitt}, D.~M. and {Hessels}, J.~W.~T. and {Kirsten}, F. and {Marcote}, B. and {Bach}, U. and {Blaauw}, R. and {Burgay}, M. and {Corongiu}, A. and {Feiler}, R. and {Gawro{\'n}ski}, M.~P. and {Giroletti}, M. and {Karuppusamy}, R. and {Keimpema}, A. and {Kharinov}, M.~A. and {Lindqvist}, M. and {Maccaferri}, G. and {Melnikov}, A. and {Mikhailov}, A. and {Ould-Boukattine}, O.~S. and {Paragi}, Z. and {Pilia}, M. and {Possenti}, A. and {Snelders}, M.~P. and {Surcis}, G. and {Trudu}, M. and {Venturi}, T. and {Vlemmings}, W. and {Wang}, N. and {Yang}, J. and {Yuan}, J.},
        title = "{Milliarcsecond Localization of the Repeating FRB 20201124A}",
      journal = {\apjl},
         year = 2022,
        month = mar,
      volume = {927},
      number = {1},
          eid = {L3},
        pages = {L3},
          doi = {10.3847/2041-8213/ac540f},
archivePrefix = {arXiv},
      eprint = {2111.01600},
 primaryClass = {astro-ph.HE},
      adsurl = {https://ui.adsabs.harvard.edu/abs/2022ApJ...927L...3N}
}

@ARTICLE{chime2023rfrb,
      author = {{Chime/Frb Collaboration} and {Andersen}, Bridget C. and {Bandura}, Kevin and {Bhardwaj}, Mohit and {Boyle}, P.~J. and {Brar}, Charanjot and {Cassanelli}, Tomas and {Chatterjee}, S. and {Chawla}, Pragya and {Cook}, Amanda M. and {Curtin}, Alice P. and {Dobbs}, Matt and {Dong}, Fengqiu Adam and {Faber}, Jakob T. and {Fandino}, Mateus and {Fonseca}, Emmanuel and {Gaensler}, B.~M. and {Giri}, Utkarsh and {Herrera-Martin}, Antonio and {Hill}, Alex S. and {Ibik}, Adaeze and {Josephy}, Alexander and {Kaczmarek}, Jane F. and {Kader}, Zarif and {Kaspi}, Victoria and {Landecker}, T.~L. and {Lanman}, Adam E. and {Lazda}, Mattias and {Leung}, Calvin and {Lin}, Hsiu-Hsien and {Masui}, Kiyoshi W. and {McKinven}, Ryan and {Mena-Parra}, Juan and {Meyers}, Bradley W. and {Michilli}, D. and {Ng}, Cherry and {Pandhi}, Ayush and {Pearlman}, Aaron B. and {Pen}, Ue-Li and {Petroff}, Emily and {Pleunis}, Ziggy and {Rafiei-Ravandi}, Masoud and {Rahman}, Mubdi and {Ransom}, Scott M. and {Renard}, Andre and {Sand}, Ketan R. and {Sanghavi}, Pranav and {Scholz}, Paul and {Shah}, Vishwangi and {Shin}, Kaitlyn and {Siegel}, Seth and {Smith}, Kendrick and {Stairs}, Ingrid and {Su}, Jianing and {Tendulkar}, Shriharsh P. and {Vanderlinde}, Keith and {Wang}, Haochen and {Wulf}, Dallas and {Zwaniga}, Andrew},
        title = "{CHIME/FRB Discovery of 25 Repeating Fast Radio Burst Sources}",
      journal = {\apj},
         year = 2023,
        month = apr,
      volume = {947},
      number = {2},
          eid = {83},
        pages = {83},
          doi = {10.3847/1538-4357/acc6c1},
archivePrefix = {arXiv},
      eprint = {2301.08762},
 primaryClass = {astro-ph.HE},
      adsurl = {https://ui.adsabs.harvard.edu/abs/2023ApJ...947...83C}
}

@ARTICLE{lidi2021,
      author = {{Li}, D. and {Wang}, P. and {Zhu}, W.~W. and {Zhang}, B. and {Zhang}, X.~X. and {Duan}, R. and {Zhang}, Y.~K. and {Feng}, Y. and {Tang}, N.~Y. and {Chatterjee}, S. and {Cordes}, J.~M. and {Cruces}, M. and {Dai}, S. and {Gajjar}, V. and {Hobbs}, G. and {Jin}, C. and {Kramer}, M. and {Lorimer}, D.~R. and {Miao}, C.~C. and {Niu}, C.~H. and {Niu}, J.~R. and {Pan}, Z.~C. and {Qian}, L. and {Spitler}, L. and {Werthimer}, D. and {Zhang}, G.~Q. and {Wang}, F.~Y. and {Xie}, X.~Y. and {Yue}, Y.~L. and {Zhang}, L. and {Zhi}, Q.~J. and {Zhu}, Y.},
        title = "{A bimodal burst energy distribution of a repeating fast radio burst source}",
      journal = {\nat},
         year = 2021,
        month = oct,
      volume = {598},
      number = {7880},
        pages = {267-271},
          doi = {10.1038/s41586-021-03878-5},
archivePrefix = {arXiv},
      eprint = {2107.08205},
 primaryClass = {astro-ph.HE},
      adsurl = {https://ui.adsabs.harvard.edu/abs/2021Natur.598..267L}
}

@ARTICLE{platts2019,
      author = {{Platts}, E. and {Weltman}, A. and {Walters}, A. and {Tendulkar}, S.~P. and {Gordin}, J.~E.~B. and {Kandhai}, S.},
        title = "{A living theory catalogue for fast radio bursts}",
      journal = {\physrep},
         year = 2019,
        month = aug,
      volume = {821},
        pages = {1-27},
          doi = {10.1016/j.physrep.2019.06.003},
archivePrefix = {arXiv},
      eprint = {1810.05836},
 primaryClass = {astro-ph.HE},
      adsurl = {https://ui.adsabs.harvard.edu/abs/2019PhR...821....1P}
}

@ARTICLE{niuc2022,
      author = {{Niu}, C. -H. and {Aggarwal}, K. and {Li}, D. and {Zhang}, X. and {Chatterjee}, S. and {Tsai}, C. -W. and {Yu}, W. and {Law}, C.~J. and {Burke-Spolaor}, S. and {Cordes}, J.~M. and {Zhang}, Y. -K. and {Ocker}, S.~K. and {Yao}, J. -M. and {Wang}, P. and {Feng}, Y. and {Niino}, Y. and {Bochenek}, C. and {Cruces}, M. and {Connor}, L. and {Jiang}, J. -A. and {Dai}, S. and {Luo}, R. and {Li}, G. -D. and {Miao}, C. -C. and {Niu}, J. -R. and {Anna-Thomas}, R. and {Sydnor}, J. and {Stern}, D. and {Wang}, W. -Y. and {Yuan}, M. and {Yue}, Y. -L. and {Zhou}, D. -J. and {Yan}, Z. and {Zhu}, W. -W. and {Zhang}, B.},
        title = "{A repeating fast radio burst associated with a persistent radio source}",
      journal = {\nat},
         year = 2022,
        month = jun,
      volume = {606},
      number = {7916},
        pages = {873-877},
          doi = {10.1038/s41586-022-04755-5},
archivePrefix = {arXiv},
      eprint = {2110.07418},
 primaryClass = {astro-ph.HE},
      adsurl = {https://ui.adsabs.harvard.edu/abs/2022Natur.606..873N}
}

@ARTICLE{LiangY2025,
       author = {{Liang}, Yi-Fang and {Li}, Ye and {Tang}, Zhen-Fan and {Yang}, Xuan and {Zhang}, Song-Bo and {Yang}, Yuan-Pei and {Wang}, Fa-Yin and {Wang}, Bao and {Xiao}, Di and {Zhao}, Qing and {Wei}, Jun-Jie and {Geng}, Jin-Jun and {Niu}, Jia-Rui and {Zhang}, Jun-Shuo and {Chen}, Guo and {Fang}, Min and {Wu}, Xue-Feng and {Dai}, Zi-Gao and {Zhu}, Wei-Wei and {Jiang}, Peng and {Zhang}, Bing},
        title = "{A Possible Periodic Rotation Measure Evolution in the Repeating FRB 20220529}",
      journal = {\apjl},
         year = 2025,
        month = nov,
       volume = {994},
       number = {1},
          eid = {L32},
        pages = {L32},
          doi = {10.3847/2041-8213/ae1d60},
archivePrefix = {arXiv},
       eprint = {2505.10463},
 primaryClass = {astro-ph.HE},
       adsurl = {https://ui.adsabs.harvard.edu/abs/2025ApJ...994L..32L}
}

@ARTICLE{stare22020,
      author = {{Bochenek}, C.~D. and {Ravi}, V. and {Belov}, K.~V. and {Hallinan}, G. and {Kocz}, J. and {Kulkarni}, S.~R. and {McKenna}, D.~L.},
        title = "{A fast radio burst associated with a Galactic magnetar}",
      journal = {\nat},
         year = 2020,
        month = nov,
      volume = {587},
      number = {7832},
        pages = {59-62},
          doi = {10.1038/s41586-020-2872-x},
archivePrefix = {arXiv},
      eprint = {2005.10828},
 primaryClass = {astro-ph.HE},
      adsurl = {https://ui.adsabs.harvard.edu/abs/2020Natur.587...59B}
}

@ARTICLE{chime2020sgr,
      author = {{CHIME/FRB Collaboration} and {Andersen}, B.~C. and {Bandura}, K.~M. and {Bhardwaj}, M. and {Bij}, A. and {Boyce}, M.~M. and {Boyle}, P.~J. and {Brar}, C. and {Cassanelli}, T. and {Chawla}, P. and {Chen}, T. and {Cliche}, J. -F. and {Cook}, A. and {Cubranic}, D. and {Curtin}, A.~P. and {Denman}, N.~T. and {Dobbs}, M. and {Dong}, F.~Q. and {Fandino}, M. and {Fonseca}, E. and {Gaensler}, B.~M. and {Giri}, U. and {Good}, D.~C. and {Halpern}, M. and {Hill}, A.~S. and {Hinshaw}, G.~F. and {H{\"o}fer}, C. and {Josephy}, A. and {Kania}, J.~W. and {Kaspi}, V.~M. and {Landecker}, T.~L. and {Leung}, C. and {Li}, D.~Z. and {Lin}, H. -H. and {Masui}, K.~W. and {McKinven}, R. and {Mena-Parra}, J. and {Merryfield}, M. and {Meyers}, B.~W. and {Michilli}, D. and {Milutinovic}, N. and {Mirhosseini}, A. and {M{\"u}nchmeyer}, M. and {Naidu}, A. and {Newburgh}, L.~B. and {Ng}, C. and {Patel}, C. and {Pen}, U. -L. and {Pinsonneault-Marotte}, T. and {Pleunis}, Z. and {Quine}, B.~M. and {Rafiei-Ravandi}, M. and {Rahman}, M. and {Ransom}, S.~M. and {Renard}, A. and {Sanghavi}, P. and {Scholz}, P. and {Shaw}, J.~R. and {Shin}, K. and {Siegel}, S.~R. and {Singh}, S. and {Smegal}, R.~J. and {Smith}, K.~M. and {Stairs}, I.~H. and {Tan}, C.~M. and {Tendulkar}, S.~P. and {Tretyakov}, I. and {Vanderlinde}, K. and {Wang}, H. and {Wulf}, D. and {Zwaniga}, A.~V.},
        title = "{A bright millisecond-duration radio burst from a Galactic magnetar}",
      journal = {\nat},
         year = 2020,
        month = nov,
      volume = {587},
      number = {7832},
        pages = {54-58},
          doi = {10.1038/s41586-020-2863-y},
archivePrefix = {arXiv},
      eprint = {2005.10324},
 primaryClass = {astro-ph.HE},
      adsurl = {https://ui.adsabs.harvard.edu/abs/2020Natur.587...54C}
}

@ARTICLE{zhang2023review,
      author = {{Zhang}, Bing},
        title = "{The physics of fast radio bursts}",
      journal = {Reviews of Modern Physics},
         year = 2023,
        month = jul,
      volume = {95},
      number = {3},
          eid = {035005},
        pages = {035005},
          doi = {10.1103/RevModPhys.95.035005},
archivePrefix = {arXiv},
      eprint = {2212.03972},
 primaryClass = {astro-ph.HE},
      adsurl = {https://ui.adsabs.harvard.edu/abs/2023RvMP...95c5005Z}
}

@ARTICLE{chime-repeaters,
  author = {{The CHIME/FRB Collaboration} and {:} and {Andersen}, B.~C. and 
	{Bandura}, K. and {Bhardwaj}, M. and {Boubel}, P. and {Boyce}, M.~M. and 
	{Boyle}, P.~J. and {Brar}, C. and {Cassanelli}, T. and {Chawla}, P. and 
	{Cubranic}, D. and {Deng}, M. and {Dobbs}, M. and {Fandino}, M. and 
	{Fonseca}, E. and {Gaensler}, B.~M. and {Gilbert}, A.~J. and 
	{Giri}, U. and {Good}, D.~C. and {Halpern}, M. and {H{\"o}fer}, C. and 
	{Hill}, A.~S. and {Hinshaw}, G. and {Josephy}, A. and {Kaspi}, V.~M. and 
	{Kothes}, R. and {Landecker}, T.~L. and {Lang}, D.~A. and {Li}, D.~Z. and 
	{Lin}, H.-H. and {Masui}, K.~W. and {Mena-Parra}, J. and {Merryfield}, M. and 
	{Mckinven}, R. and {Michilli}, D. and {Milutinovic}, N. and 
	{Naidu}, A. and {Newburgh}, L.~B. and {Ng}, C. and {Patel}, C. and 
	{Pen}, U. and {Pinsonneault-Marotte}, T. and {Pleunis}, Z. and 
	{Rafiei-Ravandi}, M. and {Rahman}, M. and {Ransom}, S.~M. and 
	{Renard}, A. and {Scholz}, P. and {Siegel}, S.~R. and {Singh}, S. and 
	{Smith}, K.~M. and {Stairs}, I.~H. and {Tendulkar}, S.~P. and 
	{Tretyakov}, I. and {Vanderlinde}, K. and {Yadav}, P. and {Zwaniga}, A.~V.
	},
    title = "{CHIME/FRB Detection of Eight New Repeating Fast Radio Burst Sources}",
  journal = {arXiv e-prints},
archivePrefix = "arXiv",
  eprint = {1908.03507},
 primaryClass = "astro-ph.HE",
     year = 2019,
    month = aug,
  adsurl = {https://ui.adsabs.harvard.edu/abs/2019arXiv190803507T}
}

@ARTICLE{katz16,
  author = {{Katz}, J.~I.},
    title = "{How Soft Gamma Repeaters Might Make Fast Radio Bursts}",
  journal = {\apj},
archivePrefix = "arXiv",
  eprint = {1512.04503},
 primaryClass = "astro-ph.HE",
     year = 2016,
    month = aug,
  volume = 826,
      eid = {226},
    pages = {226},
      doi = {10.3847/0004-637X/826/2/226},
  adsurl = {http://adsabs.harvard.edu/abs/2016ApJ...826..226K}
}

@ARTICLE{Pandhi2026,
       author = {{Pandhi}, Ayush and {Nimmo}, Kenzie and {Andrew}, Shion and {Brar}, Charanjot and {Chatterjee}, Shami and {Cook}, Amanda M. and {Curtin}, Alice and {Gaensler}, B.~M. and {Gawronski}, Marcin and {Hessels}, Jason and {Kaspi}, Victoria M. and {Khan}, Afrokk and {Kirsten}, Franz and {Lazda}, Mattias and {Leung}, Calvin and {Main}, Robert and {Masui}, Kiyoshi W. and {Mckinven}, Ryan and {Michilli}, Daniele and {Ng}, Mason and {Ould-Boukattine}, Omar and {Pearlman}, Aaron B. and {Pleunis}, Ziggy and {Pollak}, Alexander W. and {Pradeep E.~T.}, Sachin and {Puchalska}, Weronika and {Sammons}, Mawson W. and {Scholz}, Paul and {Shah}, Vishwangi and {Shin}, Kaitlyn and {Siegel}, Seth R. and {Smith}, Kendrick},
        title = "{A Steadily Declining Dispersion Measure for the Repeating Fast Radio Burst FRB 20220529A: Evidence for a Fast Radio Burst Engine Embedded in an Expanding Supernova Remnant}",
      journal = {\apjl},
         year = 2026,
        month = apr,
       volume = {1000},
       number = {2},
          eid = {L53},
        pages = {L53},
          doi = {10.3847/2041-8213/ae52f8},
archivePrefix = {arXiv},
       eprint = {2602.22309},
 primaryClass = {astro-ph.HE},
       adsurl = {https://ui.adsabs.harvard.edu/abs/2026ApJ..1000L..53P}
}

@ARTICLE{Lorimer2007,
  author = {{Lorimer}, D.~R. and {Bailes}, M. and {McLaughlin}, M.~A. and 
	{Narkevic}, D.~J. and {Crawford}, F.},
    title = "{A Bright Millisecond Radio Burst of Extragalactic Origin}",
  journal = {Science},
archivePrefix = "arXiv",
  eprint = {0709.4301},
     year = 2007,
    month = nov,
  volume = 318,
    pages = {777},
      doi = {10.1126/science.1147532},
  adsurl = {http://adsabs.harvard.edu/abs/2007Sci...318..777L}
}
